\documentclass[galaxies, review, submit, pdftex, moreauthors]{Definitions/mdpi} 

\usepackage{xspace}
\usepackage{pifont}

\firstpage{1} 
\pubvolume{1}
\issuenum{1}
\articlenumber{0}
\pubyear{2022}
\copyrightyear{2022}
\datereceived{} 
\dateaccepted{} 
\datepublished{} 
\hreflink{https://doi.org/} 
\preto{\abstractkeywords}{\nolinenumbers}

\Title{Open-source Radiative Modelling Tools for Extragalactic VHE gamma-ray Sources}

\TitleCitation{Open-source Radiative Modelling Tools for Extragalactic VHE gamma-ray Sources}

\Author{Cosimo Nigro $^{1}$\orcidA{}, Andrea Tramacere $^{2}$\orcidB{}}

\AuthorNames{Cosimo Nigro and Andrea Tramacere}

\AuthorCitation{Nigro, C.; Tramacere, A.}

\address{%
$^{1}$ \quad Institut de Física d’Altes Energies (IFAE), The Barcelona Institute of Science and Technology, Campus UAB, 08193 Bellaterra (Barcelona), Spain; cosimo.nigro@ifae.es\\
$^{2}$ \quad Department of Astronomy, University of Geneva, ch. d’Ecogia 16, CH-1290 Versoix, Switzerland; andrea.tramacere@unige.ch}

\abstract{In this review, we discuss various open-source software for modelling the broad-band emission of extragalactic sources from radio up to the highest gamma ray energies. As we provide an overview of the different tools available, we discuss the physical processes such tools implement and detail the computations they can perform. We also examine their conformity with modern good software practices. After considering the currently available software as a first generation of open-source modelling tools, we outline some desirable characteristics for the next generation.}


\newcommand{\python}{\texttt{python}\xspace}
\newcommand{\CC}{\texttt{C}\xspace}
\newcommand{\CPP}{\texttt{C++}\xspace}
\newcommand{\pip}{\texttt{pip}\xspace}
\newcommand{\conda}{\texttt{conda}\xspace}
\newcommand{\make}{\texttt{make}\xspace}
\newcommand{\github}{\texttt{GitHub}\xspace}
\newcommand{\githubpages}{\texttt{GitHub Pages}\xspace}
\newcommand{\readthedocs}{\texttt{Read the Docs}\xspace}
\newcommand{\numpy}{\texttt{NumPy}\xspace}
\newcommand{\scipy}{\texttt{SciPy}\xspace}
\newcommand{\astropy}{\texttt{astropy}\xspace}
\newcommand{\sherpa}{\texttt{sherpa}\xspace}
\newcommand{\emcee}{\texttt{emcee}\xspace}
\newcommand{\Gammapy}{\texttt{Gammapy}\xspace}
\newcommand{\naima}{\texttt{naima}\xspace}
\newcommand{\gamera}{\texttt{GAMERA}\xspace}
\newcommand{\jetset}{\texttt{JetSeT}\xspace}
\newcommand{\agnpy}{\texttt{agnpy}\xspace}
\newcommand{\bhjet}{\texttt{BHJet}\xspace}
\newcommand{\kariba}{\texttt{Kariba}\xspace}
\newcommand{\bljet}{\texttt{BlJet}\xspace}
\newcommand{\agnjet}{\texttt{AgnJet}\xspace}
\newcommand{\xspec}{\texttt{XSPEC}\xspace}
\newcommand{\flaremodel}{\texttt{FLAREMODEL}\xspace}
\providecommand{\doeprint}[1]{\href{http://ascl.net/#1}{\nolinkurl{http://ascl.net/#1}}}
\newcommand{\ergsecond}{{\rm erg}\,{\rm s}^{-1}}
\newcommand{\cmark}{\ding{51}}%
\newcommand{\xmark}{\ding{55}}%

\begin{document}

\section{Introduction}
In the last two decades, the energy window in which active galactic nuclei (AGN) and gamma-ray bursts (GRBs) can be observed has been extended towards high ($E > 100\,{\rm MeV}$) and very-high ($E > 100\,{\rm GeV}$) energies \cite{madejski_2016, dermer_2016, piron_2016, nava_2021, noda_2022}. The broad-band emission of extra-galactic sources, from radio to gamma rays, is commonly modelled with the radiative processes of non-thermal relativistic particles \cite{rybicki_1986, dermer_2009a, cerruti_2020}. This modelling approach offers the promising prospect to study astrophysical acceleration mechanisms and, ultimately, identify the sources of cosmic rays \cite{boettcher_2019}. Traditionally, once multi-wavelength (MWL) data are gathered and reduced, their interpretation is performed with closed-source software\footnote{It is difficult to establish if this software is proprietary, i.e. protected by a license and with limited distribution or simply private, i.e. accessible without a license to a small group of developers. We therefore apply the more general term \textit{closed-source}, meaning it cannot be publicly studied, changed or distributed.}. In the years, the growing amount and coverage of MWL data resulted in the production of several closed-source modelling software with increasing complexity, all inevitably engendering the issue of reproducibility of results\footnote{By reproducibility of results we mean the possibility for a user to download the software and the scripts associated to a certain publication (and possibly the computational environment e.g. in the form of a container) and re-perform the calculations in autonomy.}. Moreover, despite often implementing the same physical processes, these software were never validated against each other, and only recently a systematic comparison of their results has been publicly presented \cite{cerruti_2021}. While appreciating that these tools forged the current understanding of the emission of extra-galactic sources, we observe that their validation and the reproducibility of their results remain inevitable limitations. In the context of the forthcoming era of high-energy astrophysics, the limited accessibility of these closed-source software represent another drawback. The next generation of gamma-ray observatories, as the Cherenkov Telescope Array (CTA), will indeed provide open access to their data \cite{lamanna_2015}. Preparing for this, astrophysicists have started to develop standardised data formats \cite{deil_2017a, nigro_2021} and open-source analysis tools \cite{knodlseder_2016, deil_2017b}. The amount of MWL data that the new observatories will make available in the future makes the old closed-source modelling approach simply not sustainable, urging for it to be opened to a wider number of astrophysicists. This introduces the necessity to provide modelling tools with open-source licenses and adopting modern good software practices. In this review, we present the outset of this shift in the modelling paradigm. We describe several open-source software publicly available for interpreting the non-thermal emission of extragalactic jetted sources. We focus our attention on tools capable to describe the highest-energy emission of these sources. Our review has a strong bias towards AGN modelling since their emission up to ${\rm TeV}$ energies is long consolidated while for GRBs this represents a more recent finding \cite{nava_2021, noda_2022}. We also briefly examine the capability of the presented frameworks to model jetted sources of galactic origin (e.g. microquasars) characterised by the same radiative processes of their extragalactic counterparts.
\par
This review is thus structured. In Section~\ref{sec:physics}, we provide a quick overview of the major physical processes at play in extragalactic jetted sources, being their knowledge essential to understand what is being implemented by the different modelling tools. Section~\ref{sec:tools} presents the software publicly available, introducing the physical processes they model and their technical specifications (e.g. language in which they are implemented, resources available, etc.). For each tool, we also provide examples of its application. In Section~\ref{sec:discussion} we critically review the available tools and suggest some desirable characteristics for a future generation of open-source modelling software. We close this review with some remarks on how to establish a more reproducible process of physical modelling and interpretation.

\section{Physical Background}
\label{sec:physics}
This section is meant to provide a brief physical background to the understanding of the astrophysical sources and physical processes modelled in each software.

\subsection{Jetted extragalactic sources}
The presence of relativistic jets is ubiquitous to radio-loud AGN and GRBs (and to some galactic sources). Jets are generated in the environments surrounding accreting black holes (BH), where collimated outflows of plasma can be launched with bulk velocities close to the speed of light $c$ \cite{blandford_1977, blandford_1982}. The main differences between these two categories of extragalactic sources reside in the velocity of the jet ($\Gamma_{\rm AGN} \sim 10$, $\Gamma_{\rm GRB} \sim 100$, with $\Gamma$ bulk Lorentz factor of the outflow), in the dissipation of energy along the jet, and in the luminosity and duration of their emission.
\par
AGN are galaxies characterised by strong nuclear activity, commonly associated to the accretion onto a supermassive BH with mass $M_{\rm BH} = 10^{6} \mbox{--} 10^{9}\,M_{\odot}$ (with $M_{\odot}$ mass of the Sun). Accretion efficiently converts gravitational energy in thermal radiation, with a range of emission from infrared (IR) to X rays, and a total luminosity reaching up to $10^{46} \mbox{--} 10^{47}\,\ergsecond$ \cite{meier_2012}. The emission in radio-loud AGN extends from radio to gamma rays (up to few ${\rm TeV}$ in some cases) and is dominated by the non-thermal radiation of the plasma flowing in the jet. As small plasmoids are observed streaming along the jet in the radio band \cite{cohen_1971}, the non-thermal emission is commonly attributed to a small region of the outflow. The so-called \textit{blazars} represent the class of jetted AGN most commonly observed at the highest energies. A small viewing angle between the jet axis and the observer \citep{blandford_1978} results in a strong relativistic boosting of their non-thermal radiation. Blazars show strong flux variability, with time scales ranging from minutes to decades \citep{ulrich_1997, rieger_2019}, and high polarisation in the optical band \cite{zhang_2019}. Blazars can be divided in two classes: BL Lacs, characterised by featureless optical spectra, and flat spectrum radio quasars (FSRQs), showing significant line and thermal emission from hot gas orbiting the central BH. The photon fields produced by line and thermal emitters, located at sub-${\rm pc}$ distances from the BH, should absorb via $\gamma\gamma$ pair production the gamma-ray spectra of FSRQs. Due to the rare observations of these absorption features (1 object out of 10) \cite{costamante_2018, meyer_2019}, it is inferred that the non-thermal emission region is located mostly at $\sim{\rm pc}$ scales.
\par
GRBs are associated with the formation of a stellar-mass BH ($M_{\rm BH}\sim 10\,M_{\odot}$) following the merger of compact objects or a catastrophic star collapse. The ensuing outflow of plasma can be collimated in a jet \cite{kumar_2015}, with particles accelerated in its progressive shocks. Non-thermal emission occurs at two characteristic distances, both well under the ${\rm pc}$ scale, characterising two different phases of the emission (\textit{prompt} and \textit{afterglow}) \cite{piron_2016}. The afterglow phase is characterised by the interaction of the shocked material with the external (or circumstellar) medium, that results in the gradual decrease in the bulk Lorentz factor and in a fainting emission in some cases detectable for days. GRB luminosities can achieve values of $10^{50} \mbox{--} 10^{52}\,\ergsecond$.

\subsection{Physical processes}
The non-thermal emission spectrum of different types of extragalactic sources can be described by a power law (PL) over a broad range of photon energies. This PL of photons is the imprint of the PL energy distribution of the radiating particles, result of first and second order Fermi acceleration processes \cite{matthews_2020}. It is commonly assumed that particles are accelerated and radiate in a finite region of the jet, that we will refer to as emission region. Being the radiation emitted in a finite region moving at relativistic speeds, its intensity is relativistically boosted by a factor depending on $\Gamma$ and on geometry. If a simple spherical plasmoid is considered, as in jetted AGN, the observed energy flux, $\nu F_{\nu}\,[{\rm erg}\,{\rm cm}^{-2}\,{\rm s}^{-1}]$, is boosted by the fourth power of the Doppler factor $\delta_D = \frac{1}{\Gamma(1 - \beta\cos\theta)}$, where $\beta$ is the velocity of the outflow and $\theta$ the observer's viewing angle. For GRBs, being the emission region extended to the whole jet section, the beaming pattern can be more complex \cite[see Sect. 2]{kumar_2015}.
\par
The spectral energy distribution (SED) of radio-loud AGN shows two main components: a low-energy one, peaking in the IR to X-ray band, commonly attributed to synchrotron radiation of relativistic electrons and positrons, and a high-energy component, peaking in gamma rays, that can be either \textit{leptonic} or \textit{hadronic} in origin. In leptonic models, the high-energy emission is due to inverse Compton (IC) scattering by ${\rm e}^{\pm}$ \cite{blandford_1979} of target photon fields internal or external to the emission region. In what is commonly referred to as \textit{synchrotron self-Compton} (SSC) scattering, the target photon field is internal to emission region and is provided by the synchrotron radiation of the very same accelerated ${\rm e}^{\pm}$ \cite{jones_1974}. In the so called \textit{external Compton} (EC) scenario, the photon fields target for Compton scattering can be provided either by the AGN line and thermal emitters: accretion disk, broad line region (BLR), dust torus (DT) \cite{sikora_1994, dermer_2002, sikora_2002}, or by the synchrotron emission of other components of the jet \cite{tavecchio_2008, macdonald_2015}, or even by the cosmic microwave background (CMB) \cite{tavecchio_2000}. In hadronic models, on the other hand, the high-energy emission is explained with the radiative processes of the secondaries originated in ${\rm p}{\rm p}$ or ${\rm p}\gamma$ interactions \cite{cerruti_2021, boettcher_2013}. The soft photon fields target for IC or ${\rm p}\gamma$ interactions can also produce absorption of the high-energy radiation via $\gamma\gamma$ pair production. The same absorption can occur on the extragalactic background light (EBL) while the photons travel to Earth \cite{cooray_2016}.
\par
Given their very recent observations at the highest energies \cite{nava_2021}, the mechanisms of the broad-band emission of GRBs are still under discussion. While some authors accommodate the whole MWL emission, up to ${\rm TeV}$ energies, with synchrotron radiation \cite{hess_grb190829}, others observed the presence of a second component at the highest energies, attributed, as in blazars, to SSC \cite{magic_grb190114c}. Hadronic models have also been suggested to accommodate the highest energy emission from GRBs \cite{razzaque_2010}.
\par
Assuming a particular physical scenario and tacking into account the corresponding radiative processes, one can fit the observed MWL SED (see e.g. Figures~\ref{fig:naima_grb}, \ref{fig:gamera_oj_287}, \ref{fig:gamera_pks_1830}) and hence infer the underlying particle energy distribution. The parameter space for these models is often degenarate, with changes in different parameters giving rise to similar patterns in the observed fluxes. Time-resolved SED modelling, properly taking into account the interplay between particles acceleration, cooling via radiative processes, and change of physical conditions in the emission region, constitutes a powerful tool to identify the physical mechanism or parameter responsible for a given observed emission. To obtain broad-band spectra at different times (see e.g. Figure~\ref{fig:exp_vs_no_exp}) it is necessary to solve a differential equation regulating the time evolution of the underlying particle energy distribution \cite[see e.g.][Eq. 7]{cerruti_2020}.

\section{Open-source modelling tools}
\label{sec:tools}
In this section we provide an overview of the open-source software publicly available to model the broad-band emission of extragalactic jetted sources. We detail both their theoretical background, list the physical processes they model, and offer an overview of their resources and usage. In the following, we identify with the term \textit{validation} the numerical comparison of the results produced by a given tool with the output of another software or the reproduction of results from the literature.

\subsection{\naima}

\begin{figure}[H]
    \centering
    \includegraphics[width=0.95\textwidth,angle=0]{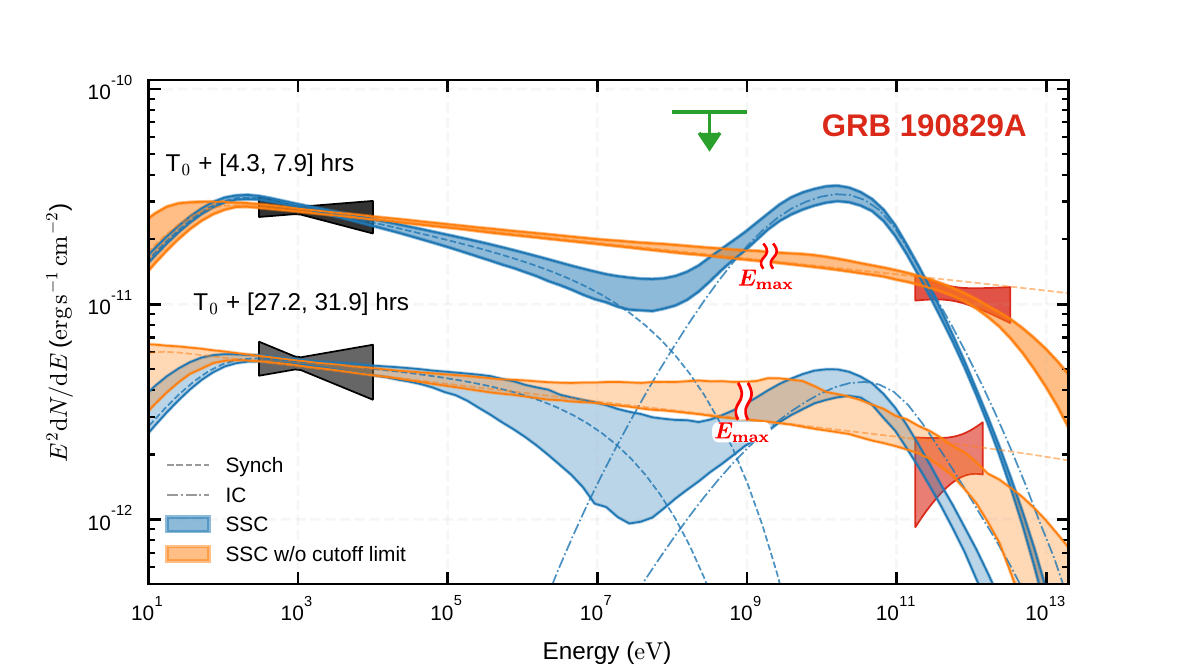}
    \caption{Synchrotron (orange) and SSC (blue) models for the emission GRB~190829A computed with \naima. The shaded area represents the $68\%$ confidence interval obtained from the MCMC fitting. Flux measurements from \textit{Swift}-XRT (black band), \textit{Fermi}-LAT (green upper limit) and H.E.S.S. (red band) are also displayed. Figure from \cite{hess_grb_190829a_2021}, reproduced with permission of The American Association for the Advancement of Science.}
    \label{fig:naima_grb}
\end{figure}

\naima \cite{zabalza_2015} is a \python package designed to infer the non-thermal particle distributions underlying an observed broad-band photon spectrum. \naima represented the first \python package modelling non-thermal radiative processes made publicly available. It is built entirely in the ecosystem formed by \numpy \cite{numpy_2020}, \scipy \cite{scipy_2020}, and \astropy \cite{astropy_paper_I, astropy_paper_II} and is one of the packages affiliated with the \astropy project\footnote{\url{https://www.astropy.org/affiliated/\#affiliated-packages}}. The packages forming this ecosystem provide the foundations on which an increasing number of tools for astrophysics are being built. For the calculation of the radiative processes, \naima relies on a numerical approach: the observed emission is computed by integrating the analytical functions representing or approximating a given emission process with the particle energy distribution. Both non-thermal leptons and hadrons distributions can be considered. \naima does not implement time evolution of the particle energy distributions, but can accept as input an arbitrary energy distribution (such that, for example, the result of a time evolution computed with another software can be considered). \naima implements both leptonic and hadronic radiative processes. Synchrotron radiation, IC on isotropic photon fields, and non-thermal Bremsstrahlung radiation are available for electrons. For protons, the photon spectrum produced by the decay of the neutral pion result of ${\rm p}{\rm p}$ interactions can be computed, though the spectrum of the secondary particles is not computed. The only $\gamma\gamma$ absorption considered is the one on the EBL, following the model of \cite{dominguez_2011}. The code assumes co-moving densities of particles therefore it is inadequate to describe extragalactic jetted sources such as blazars or GRBs, in which typically the emission region moves at relativistic speed against a target (photon fields or other particles). The beaming pattern due to the relativistic motion of the emission region is also not computed and it has to be manually calculated by the user. There is no option to consider multiple emission regions. Being the inference of the particle distribution underlying one or more radiative processes the main objective of the package, routines for flux points handling and SED fitting are provided. \naima offers a wrapper to import its radiative models in \sherpa \cite{freeman_2001, doe_2007}, allowing the user to use \sherpa's data handling and fitting capabilities. \Gammapy \cite{deil_2017b}, a python package for the analysis of gamma-ray data, includes a wrapper to the \naima radiative models in its own source code. Alternatively, a Markov chain Monte Carlo (MCMC) fit can be performed interfacing \naima with \emcee \cite{foreman_2013}, as illustrated in \cite{zabalza_2015} and in other examples in the documentation. No validation of the radiative processes implementated in the package is provided in the documentation or in \cite{zabalza_2015}.
\par
\naima has mostly been used to model galactic sources, especially supernova remnants or pulsar wind nebulae (PWN). Nonetheless, for the simplest case of an emission region with a simple geometry and moving at relativistic speed, the package can be adapted to model jetted extragalactic sources (computing a flux and boosting it a posteriori). This approach was used for Mkr~421 in \cite{magic_mrk_421_2021} and for GRB~190829A in \cite{hess_grb_190829a_2021}. The scripts using \naima for the interpretation of GRB~190829A, illustrated in Figure~\ref{fig:naima_grb}, are also available on \github\footnote{\url{https://github.com/Carlor87/GRBmodelling}}. In this case, a spherical shell emission region is considered, with electrons accelerated by a forward shock swiping material from a stellar wind or from the interstellar medium (ISM). Since \naima does not perform temporal evolution, a broken PL with an exponential cut-off is considered for the electron distribution.
\par
\naima's developement is hosted on \github\footnote{\url{https://github.com/zblz/naima}}, where 8 contributors are listed. It follows modern good software practices adopted by other \python packages: it provides a documentation hosted on \readthedocs\footnote{\url{https://naima.readthedocs.io/en/latest/}} and includes a test suite part of a Continuous Integration (CI) system. \naima can be installed via \pip and \conda. \cite{zabalza_2015} constitutes the only reference publication for the package.

\subsection{\gamera}

\begin{figure}[H]
    \centering
    \begin{tabular}{cc}
    \includegraphics[width=.50\linewidth,angle=-0]{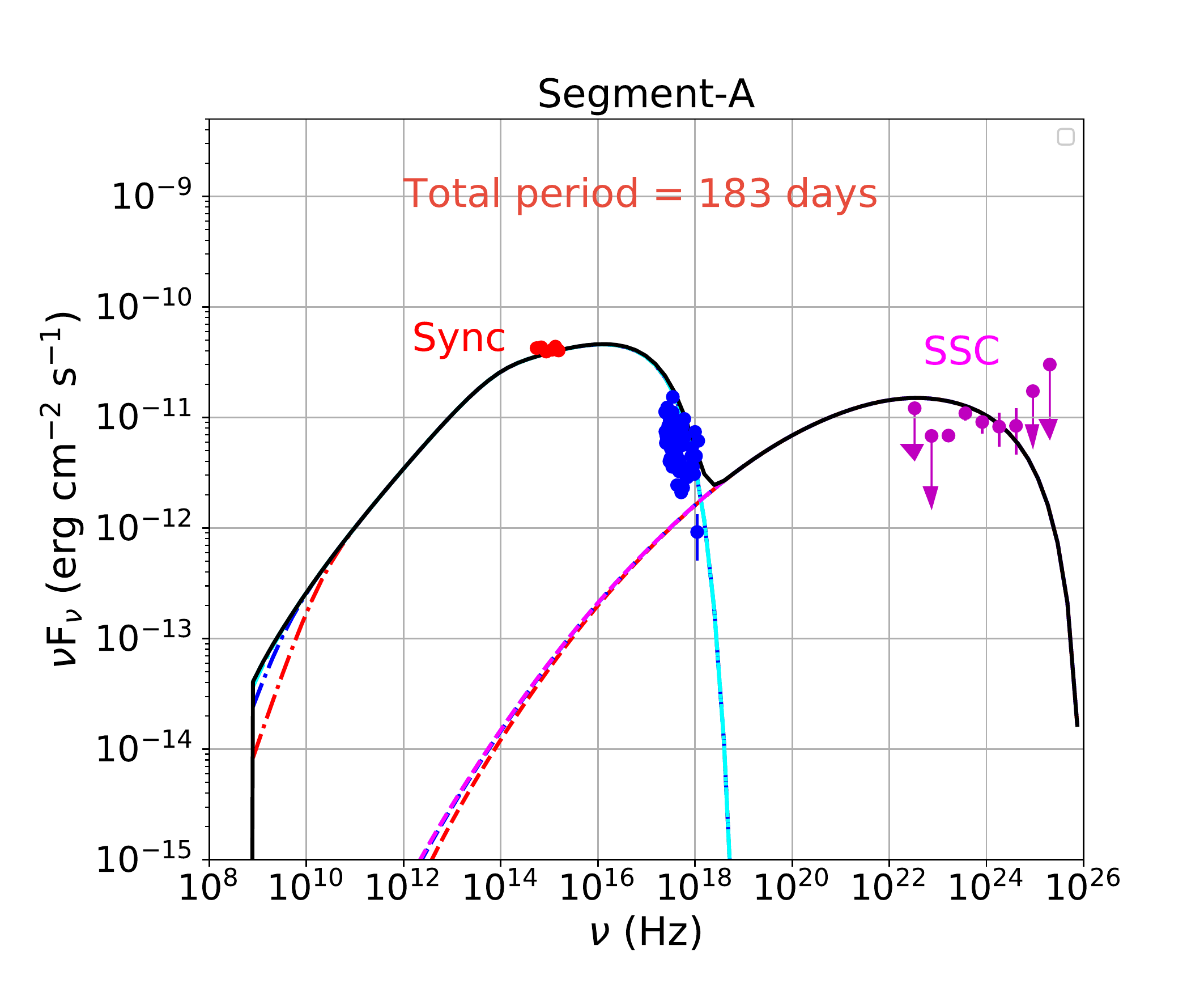}&  
    \includegraphics[width=.50\linewidth,angle=-0]{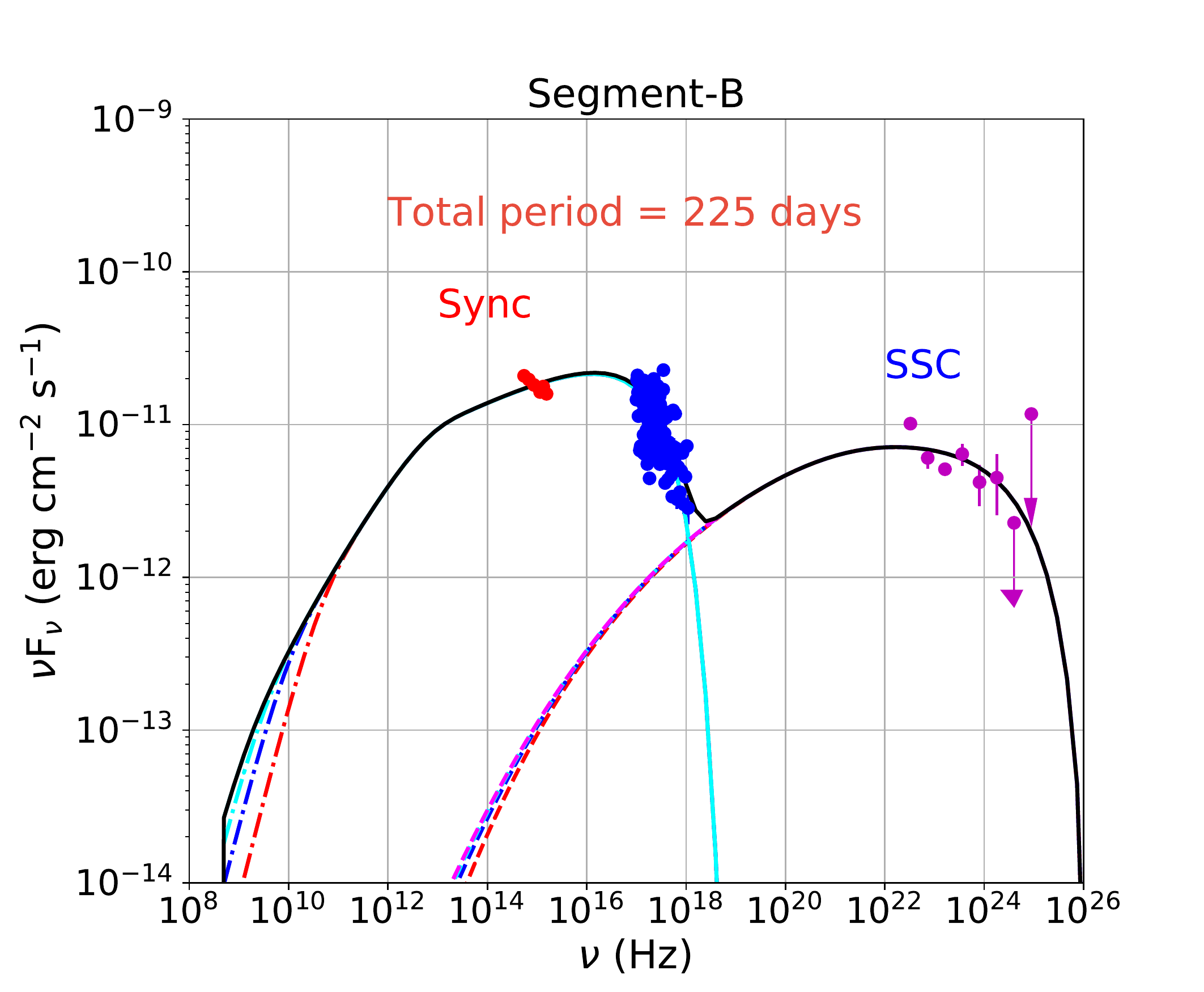}\\
    \end{tabular}
    \caption{Models for the broad-band emission from radio to gamma rays of OJ~287, for two datasets observed during the years 2017--2020. The dotted-dashed and dashed line represent synchrotron and SSC emission computed with \gamera, the different colors represent different times of the electrons distribution evolution. Optical / UV, X-ray, and gamma-ray flux measurements are displayed with red, blue, and magenta markers, respectively. Figure from \cite{prince_2021}, reproduced with permission of Astronomy and Astrophysics.}
    \label{fig:gamera_oj_287}
\end{figure}

\begin{figure}[h]
    \centering
    \includegraphics[scale=0.28]{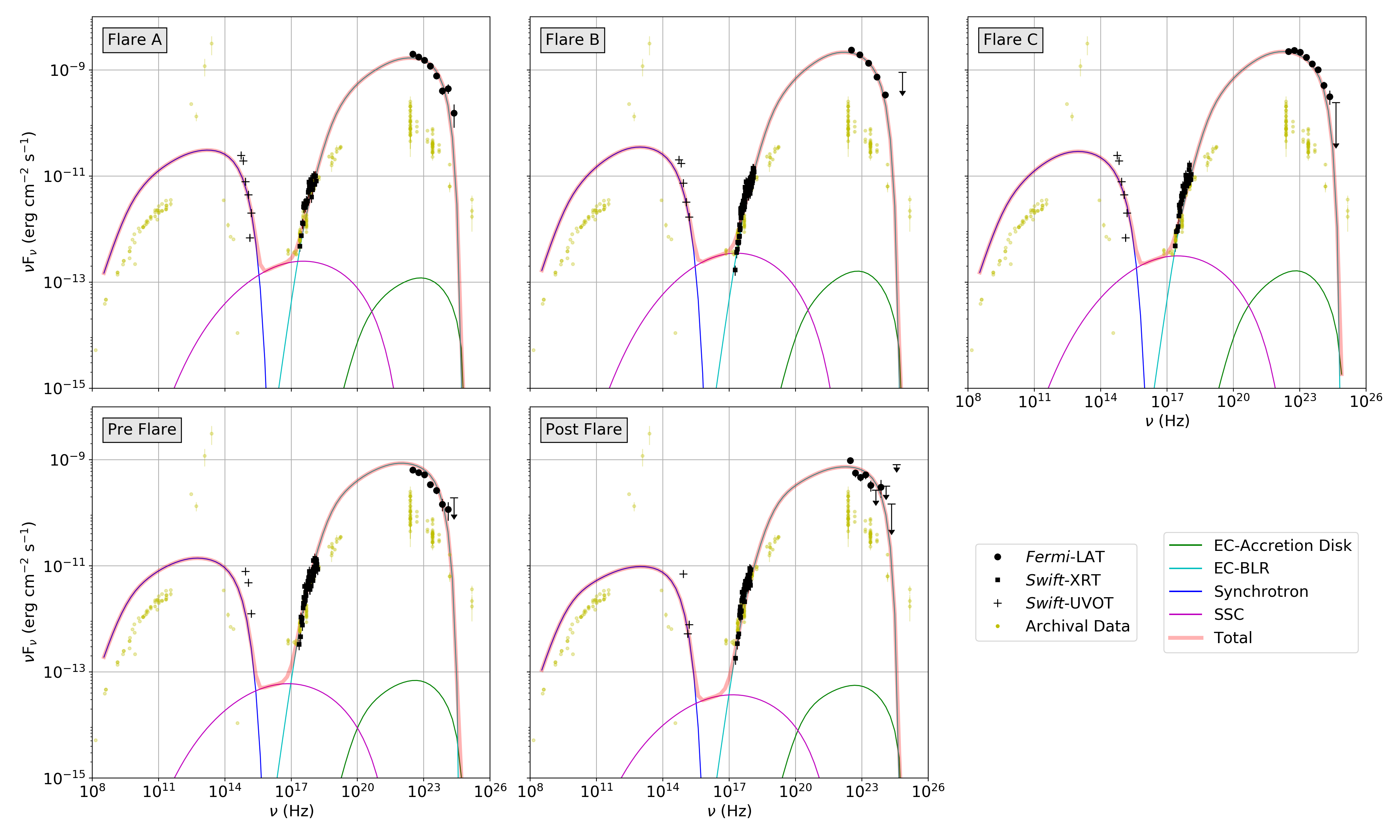}
    \caption{MWL emission of PKS~1830-211 in different flux states, modelled with \gamera. The high-energy component of the FSRQ emission is described with a combination of SSC and EC scattering on anisotropic photon fields: disk (green line) and BLR (blue line), dominating. Figure from \cite{abhir_2021}, reproduced with permission of the authors and IOPPublishing.}
    \label{fig:gamera_pks_1830}
\end{figure}

\gamera \cite{hahn_2015} is a \CPP library providing a modular approach to the modelling of the emission of different types of sources, along with some tools for population studies. \gamera offers a \python wrapper, \texttt{gappa}, returning the result of \gamera's computations as \numpy arrays. As \naima, \gamera relies on a numerical approach for the radiative processes computation. Non-thermal electron and proton distribution can be considered and can be evolved in time. It is possible to evolve the particle spectra considering cooling via all the radiative processes implemented in the package, which are the same available in \naima (synchrotron, IC, Bremmsstrahlung for ${\rm e}^{\pm}$ and decay of $\pi_0$ from ${\rm p}{\rm p}$ interactions). The numerical solution of the particle transport equation is based on an algorithm that interprets the transport as an advective flow in energy space and solves it using a donor-cell advection algorithm. For constant energy losses and no particle escape, \gamera offers the possibility to use a semi-analytical method, providing a faster computation. For Compton scatterinng, the full angular dependency of the Compton cross section is considered, allowing to model the scattering of anisotropic photon fields by anisotropic electrons. The effect of $\gamma\gamma$ absorption on anisotropic photon fields is also modelled by the library. Several interacting emission regions can be considered. No validation for the radiative processes computation is provided in \citep{hahn_2015}; though the pulsar wind nebula model in \cite{torres_2014} is reproduced, no numerical comparison of the SEDs is provided. \gamera does not directly implement data handling, nor provides wrappers to other fitting packages. A \python script for SED fitting is available in the documentation, but the model for fitting has to be manually modified by the user. \gamera implements routines for flux point\footnote{The energy flux $[{\rm erg}\,{\rm cm}^{-2}\,{\rm s}^{-1}]$ measured by an instrument in a given energy bin.} simulation: once an instrument response function (IRF) is provided, the observed flux corresponding to a specific radiative model can be obtained. Utilities representing the first steps in a population studies are provided with the library. In \cite{hahn_2015} it is shown how to generate a population of young PWN in the Galaxy, though it is left to the user to compute their total emission. 
\par
\gamera has been mostly employed for AGN modelling \cite{abhir_2021, prince_2021, matthews_2021}. In Figure~\ref{fig:gamera_oj_287} we show the model obtained with \gamera fitting the MWL emission of the BL Lac OJ~287 observed during 2017--2020 \cite{prince_2021}. The emission corresponds to an electron distribution with an injected log-parabolic spectrum cooling via synchrotron and IC radiation. In Figure~\ref{fig:gamera_pks_1830} we show instead the application of \gamera to model the highest energy component of the FSRQ PKS~1830-211 with EC scattering on photons produced by the BLR (anisotropic target photon field).
\par
\gamera's development is hosted on \github\footnote{\url{https://github.com/libgamera/GAMERA}}, where 5 contributors are listed; its documentation is hosted online\footnote{\url{https://libgamera.github.io/GAMERA/docs/main\_page.html}}. No unit tests and no CI are setup. The \CPP library is not distributed with any package system and has to be manually downloaded from \github and built with \make. The \python wrapper is not available via a standard package manager (\pip, \conda), a static library has to be manually built by the user and appended to the search path for modules in each \python script.

\subsection{\jetset}

\jetset \cite{tramacere_2009, tramacere_2011, jetset_2020} is an open-source \CC/\python framework to reproduce radiative and acceleration processes acting in extragalactic jets and galactic objects (beamed and unbeamed). Both static and time-dependent modelling are implemented, allowing the user to fit the numerical models to observed data. \jetset allows defining several leptonic radiative scenarios: synchrotron, SSC, EC on disk, BLR and DT photon fields, EC on the CMB. It also computes the $\gamma\gamma$ absorption on the EBL models of \cite{franceschini_2008, finke_2010, dominguez_2011}. Moreover, \jetset models hadronic ${\rm p}{\rm p}$ emission, considering $\gamma$ from $\pi^0$ decay, and also the radiation from the secondaries of charged pions (evolved to equilibrium). Neutrinos spectra result of the decay of these secondaries can also be estimated. \jetset incorporates template models, e.g. for the host galaxy and the \textit{big blue bump} (BBB) feature produced by the disk. The code implements a self-consistent temporal evolution of the plasma under the effect of radiative and adiabatic cooling, and both first and second order (stochastic) acceleration processes. \jetset provides tools to handle observed data such as grouping, definition of data sets, handling of upper limits and time ranges. All the data sets, the output tables, and the produced SEDs can be returned as \astropy tables with units. The model fitting can be assisted by a pre-fit stage in which a phenomenological characterisation of the SED is fitted to the data (via power-law and log-polynomial fit). The derived parameters, such as spectral indices, curvatures, peak fluxes and frequencies, are used to constrain the parameter space of the synchrotron and SSC / EC scenarios. These constraints are taken into account in the successive fit stage, with the proper physical radiative models. From an implementation point of view, \jetset is fully object-oriented, with both inheritance and composition, and provides a broad range of models, implemented inheriting from the \texttt{BaseModel} class. Models can be combined together, using the \texttt{FitModel} class from the \texttt{model\_manager} module, and then plugged to a minimiser for fitting. The main type of models are: 
\begin{itemize}
	\item {\it numerical models}:  
    \begin{itemize}
        \item{} \texttt{Jet} class, handling both leptonic and hadronic (${\rm p}{\rm p}$) emission for extragalactic jetted objects, and the \texttt{JetTimeEvol} to perform temporal evolution of a leptonic plasma,
        \item{} \texttt{GalacticBeamed} class for galactic jetted objects,
        \item{} \texttt{GalacticUnbeamed} class for galactic objects without jets, such as PWN and SNR;
    \end{itemize}
	\item {\it analytical models}: handling the phenomenological models (e.g. power-law or log-polynomial models) used for the pre-fit stage. They can be additionally used to define user-defined analytical models to plug, via the model manager, to the fitting routines;
	\item {\it template models}: used to reproduce template of the galaxy emission or of the BBB, and also used for the computation of the absorption on the EBL (with a dependency on redshift and energy).
\end{itemize}

\begin{figure}[!h]
    \centering
    \begin{tabular}{c}
    \includegraphics[width=.8\linewidth,angle=-0]{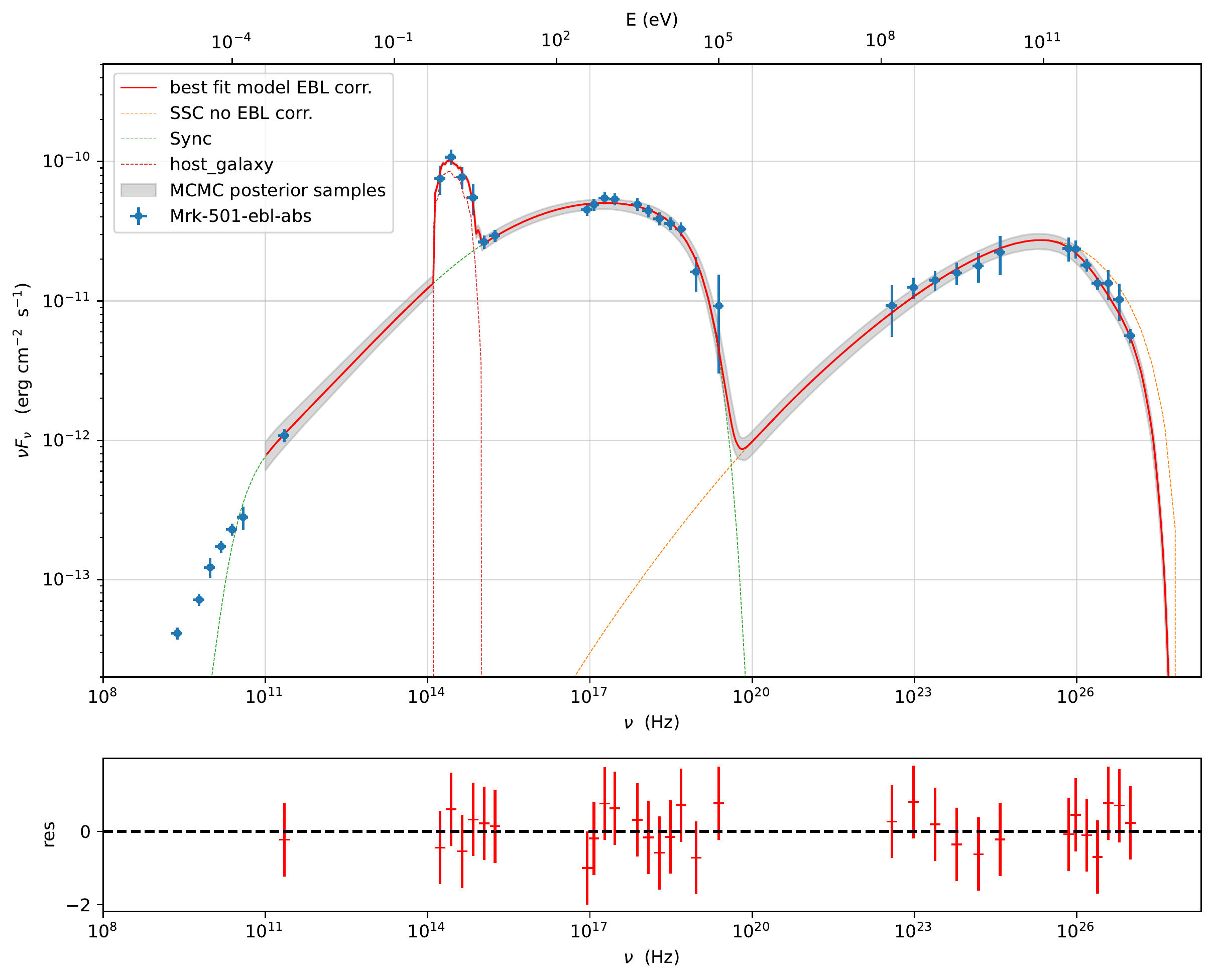}\\
    \hspace{2.em}
    \includegraphics[width=.8\linewidth,angle=-0]{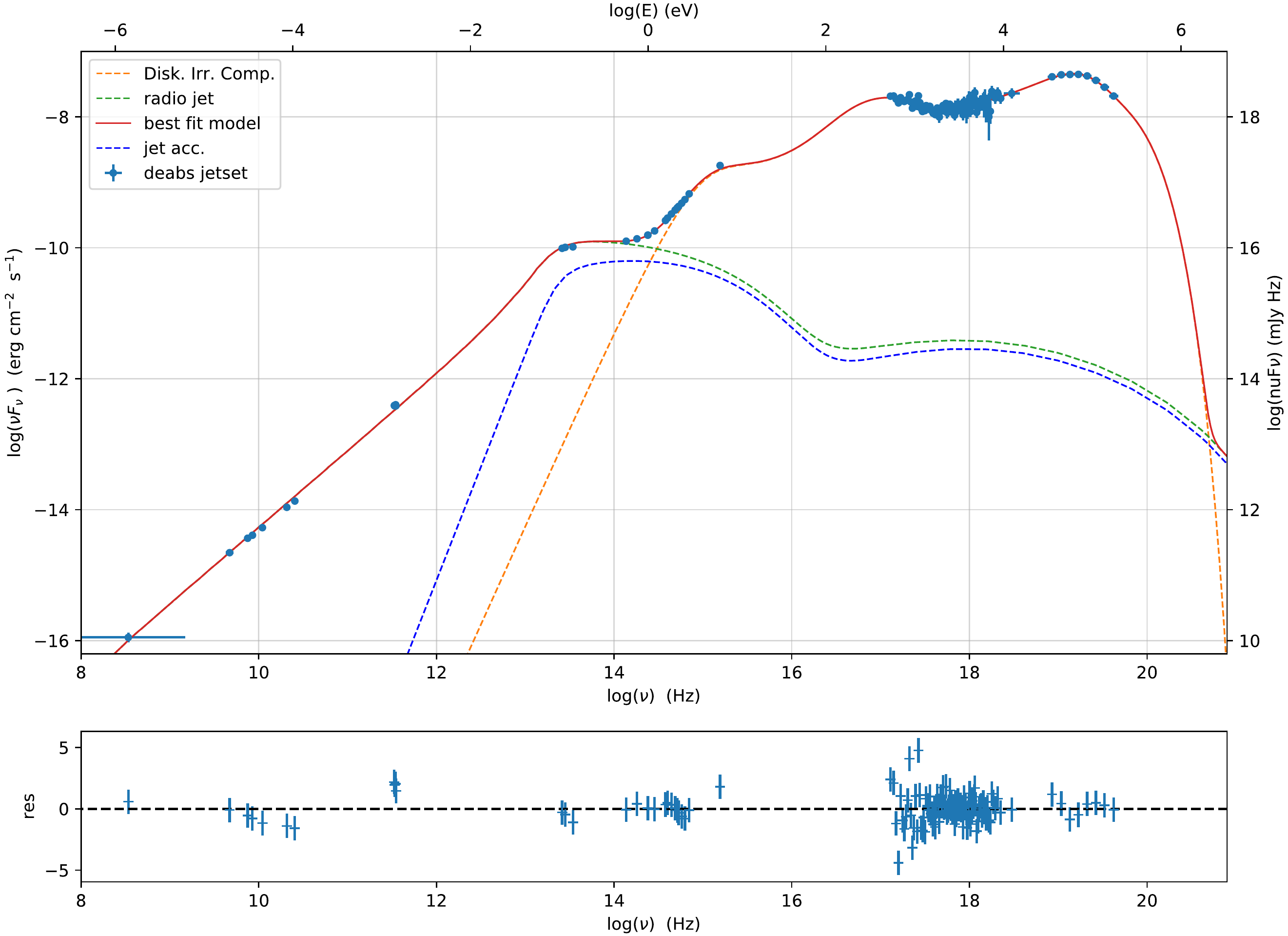}\\
    \end{tabular}
	\caption{Fitting MWL SEDs with \jetset. {\it Top panel}: best-fit of the Mrk~501 SED considering synchrotron and SSC emission, the galaxy template and the absorption on the EBL. The gray band illustrates the MCMC model posterior samples. Figure adapted from the \jetset documentation. {\it Bottom panel}: best-fit model of the MWL emission of the microquasars  MAXI~J1820+070. The dashed lines represent the individual components, the red line their sum. Figure from \cite{rodi_2021}.}
    \label{fig:jetset_fit}
\end{figure}

The parameters of the models are handled by a dedicated class which implements, via composition and inheritance, complex and flexible features. The parameters wrap \astropy \text{Quantities}, for easy interface with other \astropy-based packages. Parameters can be linked via mathematical expression, both within the same model or among different models. For example, one can: define the magnetic field as function of the blob size and position across the jet; the BLR size as a function of the disk luminosity; or set an analytical dependency between the low and high energy indexes of a broken PL particle distribution. Dedicated classes handle both frequentist and bayesian model fitting (see Figure~\ref{fig:jetset_fit}, top panel). The frequentist model fitting class implements plugin to \texttt{iminuit} \cite{iminuit} and to the \scipy least square bound implementation. The bayesian model fitting can instead be performed using a MCMC sampler with a plugin to \emcee \cite{foreman_2013}. Best fit SEDs and parameters, including MCMC results, can be stored to file. A plugin to use \sherpa and \Gammapy is also implemented. The temporal evolution of the leptonic plasma is implemented in the \texttt{JetTimeEvol} class. To follow the evolution of the particle distribution, \jetset proceeds through the numerical solution of a kinetic equation based on the the quasi-linear approximation with the inclusion of a momentum diffusion term in \citep{ramaty_1979, becker_2006}. The numerical solution of the Fokker-Planck equation is obtained using the same approach as \cite{tramacere_2011}, which is based on the method proposed by \cite{chang_1970} and \cite{park_1996}. The temporal evolution can connect together more than one region, allowing to simulate the acceleration and radiative regions separately, injecting the particles from the acceleration to the radiative region. The code allows to store particle distributions, SEDs, and light curves (with a user specified sampling) and to convolve the light curves with the light crossing time through the emission region. Each defined model, including the models with temporal evolution, can be saved using \python's pickling mechanism.

\begin{figure}[H]
    \centering
    \begin{tabular}{c}
    \includegraphics[width=.75\linewidth,angle=-0]{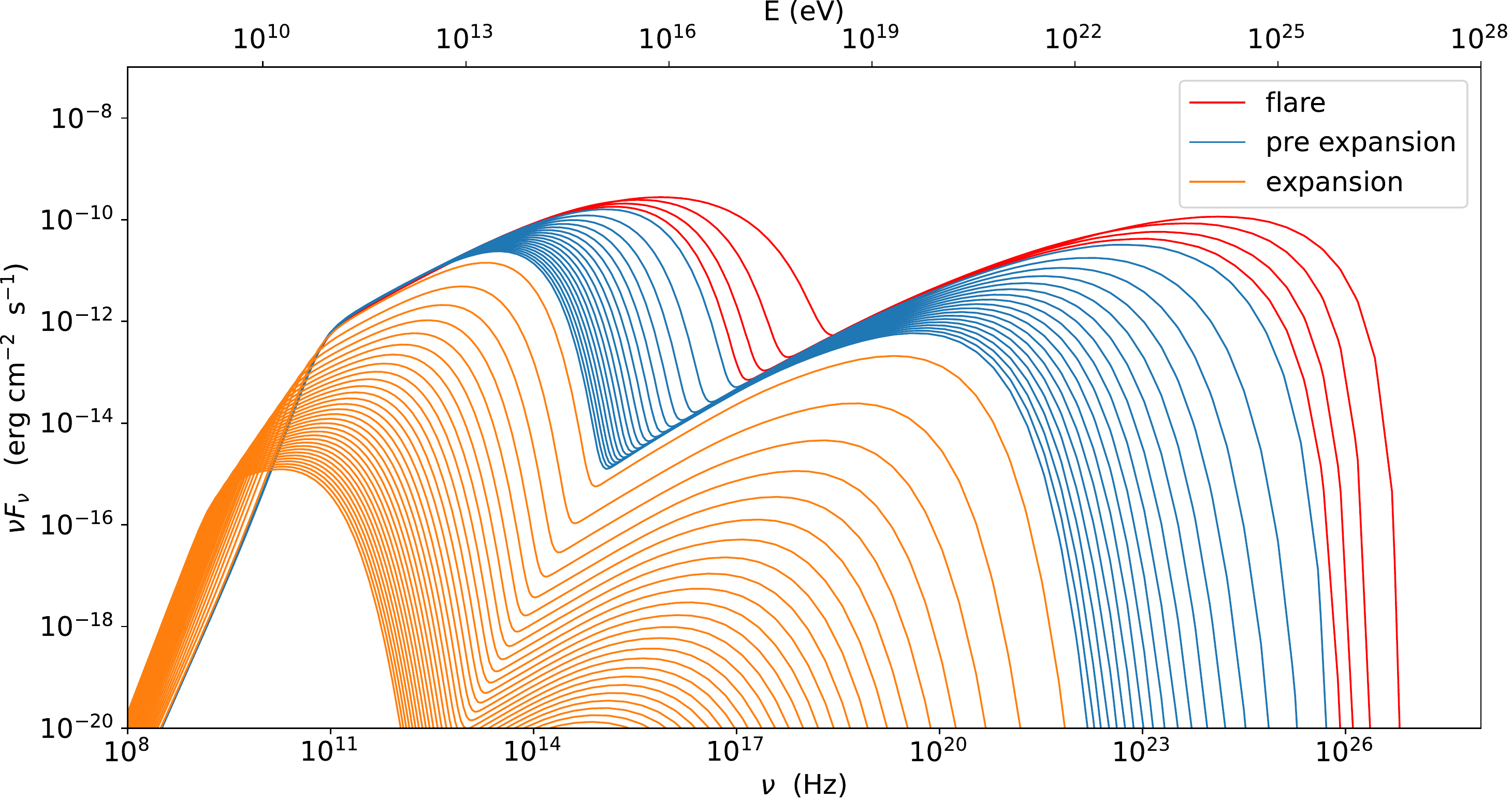}\\   
    \includegraphics[width=.75\linewidth,angle=-0]{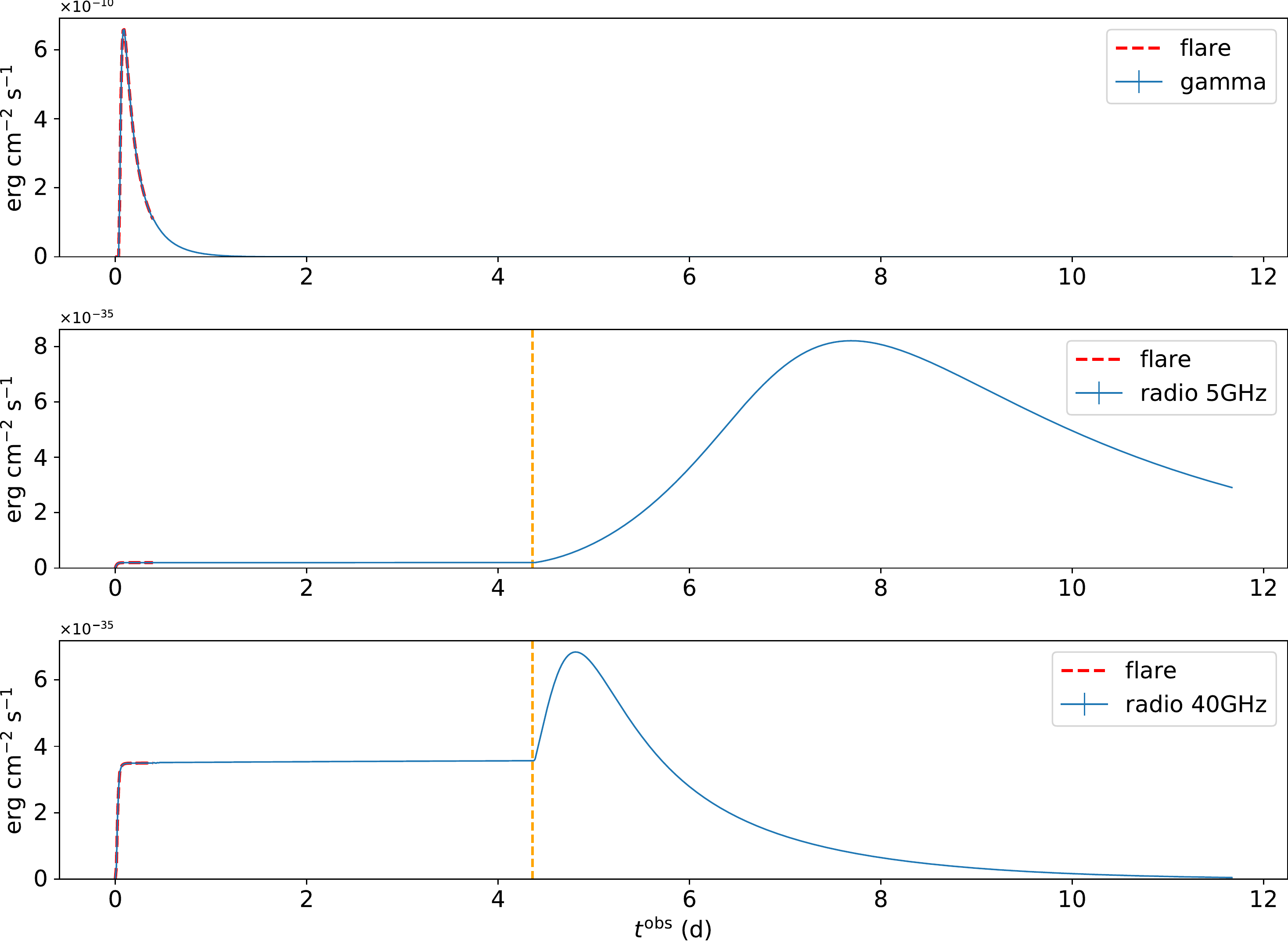}\\
    \end{tabular}
    \caption{{\it Top panel}: time-resolved SEDs, computed with \jetset, for a flaring stage (red lines) followed by a pre-expanding stage (blue lines) and an adiabatic expansion stage (orange lines) with $\beta_{\rm exp}=0.1$. The three {\it bottom panels} show the corresponding light curves at high energies, and in the radio at $5$ and $40\,{\rm GHz}$. The red dashed lines mark the light-curve segment belonging to the flaring stage and the orange vertical dashed lines mark the beginning of the expansion, the orange line marks the expansion stage. Figure adapted from \cite{tramacere_2022}.}
    \label{fig:exp_vs_no_exp}
\end{figure}

\jetset has been extensively used for modelling and fitting radiative emission in blazars both for BL~Lacs and FSRQs. Temporal evolution capabilities have been used in a recent work \cite{tramacere_2022} to simulate the impact of the adiabatic expansion on radio to gamma-ray delays. In the model in Figure~\ref{fig:exp_vs_no_exp} particles are initially injected and accelerated in an acceleration region, where they undergo both acceleration and cooling. They then diffuse towards a radiative region, where only radiative losses and adiabatic expansion take place. The effect of the expansion, leading to a decrease in the magnetic field, can be observed both in the SEDs (top panel), showing a shift of the synchrotron self-absorption frequency, and in the light curves (three bottom panels), showing the delays observed between gamma-ray and radio flares. The flexibility of the code allows building complex and flexible user defined models, or plugins, as demonstrated for the microquasars MAXI~J1820+070 during the 2018 outburst \cite{rodi_2021}. The model is composed of an irradiated disk with a Compton hump and a leptonic jet with an acceleration region and a synchrotron-dominated cooling region. Figure~\ref{fig:jetset_fit} (bottom panel) illustrates the best fit SED for this scenario obtained with \jetset with the MWL data in \cite{rodi_2021}.
\par
\jetset is hosted on \github\footnote{\url{https://github.com/andreatramacere/jetset}} and the  documentation is hosted on \readthedocs\footnote{\url{https://jetset.readthedocs.io/en/latest/}}. Continuous integration (CI) and continuous deployment (CD) are performed by GitHub Actions. Test suites are performed for each new release, that are available via \conda and \pip. Pre-releases, for source \conda and \pip, are hosted on \github\footnote{\url{https://github.com/andreatramacere/jetset/releases}}, and documented on \githubpages\footnote{\url{https://andreatramacere.github.io/jetsetdoc/html/index.html}}. Pre-releases can be easily installed using the script in the \texttt{jetset-installer}\footnote{\url{https://github.com/andreatramacere/jetset-installer}} repository.

\subsection{\agnpy}

\begin{figure}
    \centering
    \begin{tabular}{l}
    \hspace{-5.5em}
    \includegraphics[width=1.15\linewidth,angle=-0]{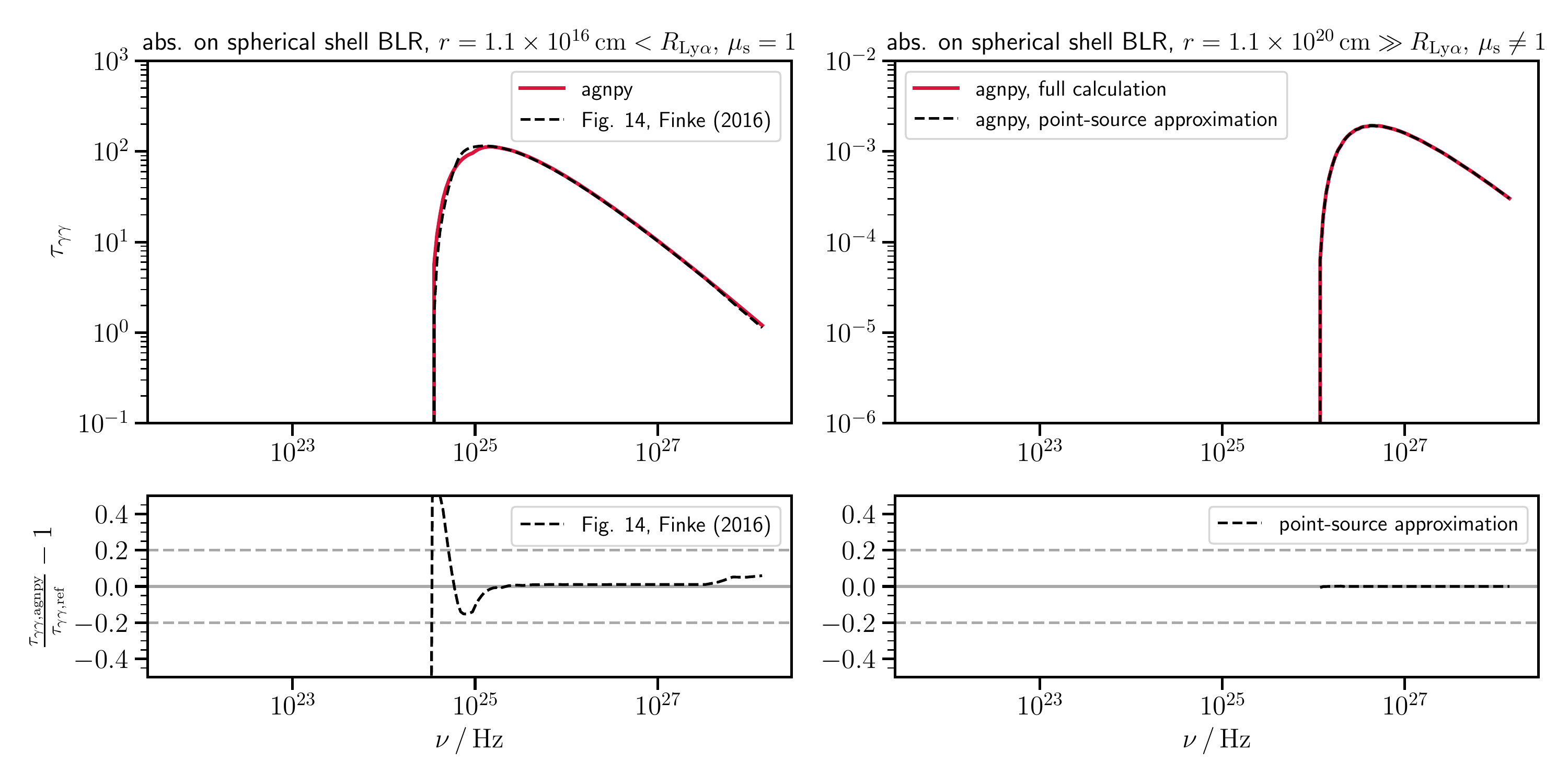}
    \end{tabular}
    \caption{$\gamma\gamma$ opacity for absorption on the BLR photon field computed with \agnpy. \textit{Left panel}: validation of \agnpy result against the literature for a small viewing angle ($\theta_{\rm s} = 0^{\circ}$, blazar case). \textit{Right panel}: internal cross-check approximating, for large distances from the BH, the BLR as a monochromatic point source at the BH position. A non-null viewing angle is considered in this case ($\theta_{\rm s} = 20^{\circ}$, radio-loud AGN case). Figure from \cite{nigro_2022}, reproduced with permission of Astronomy and Astrophysics.}
    \label{fig:blr_absorption_agnpy}
\end{figure}

\begin{figure}
    \centering
    \includegraphics[width=0.7\textwidth,angle=0]{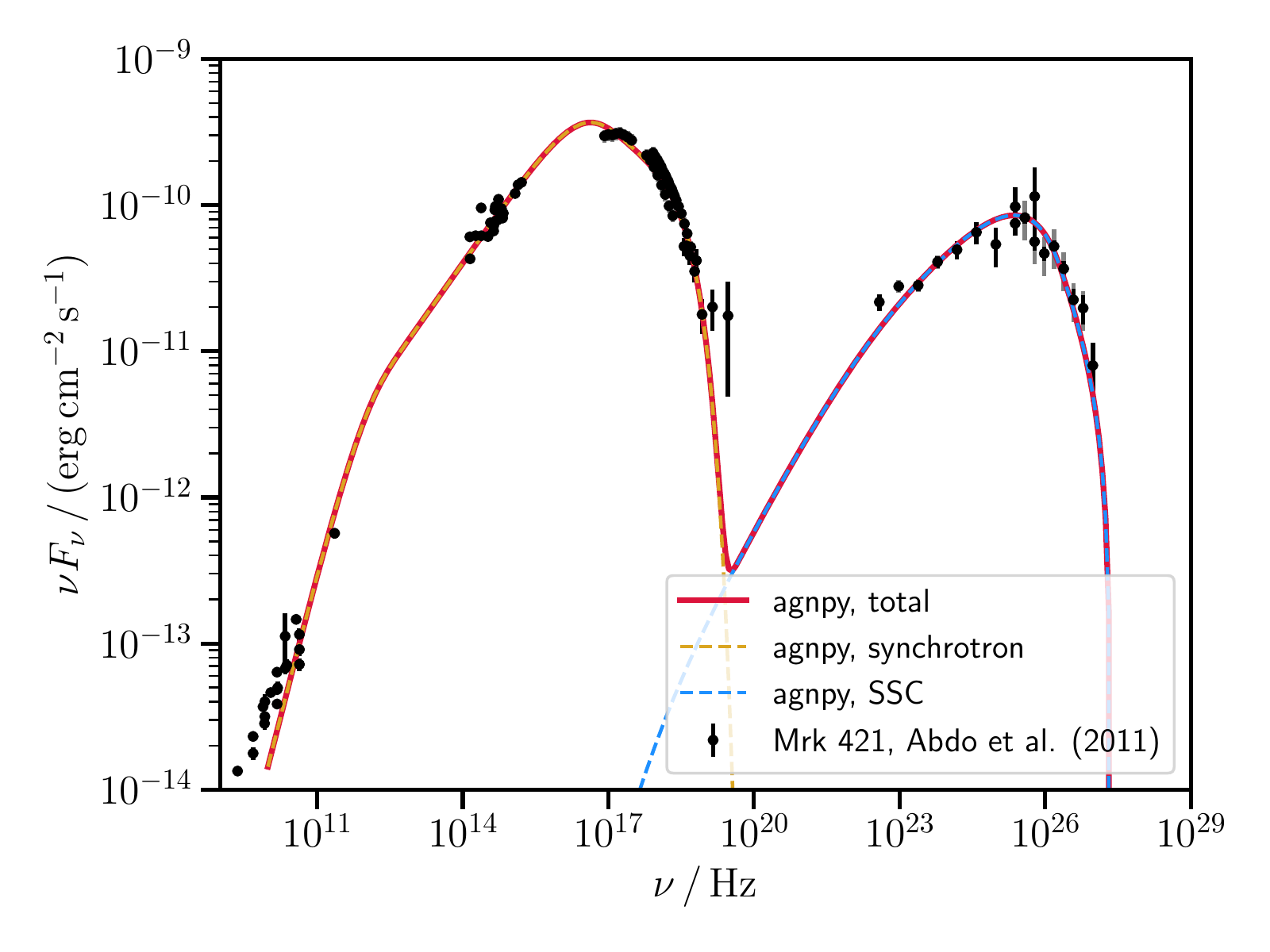}
    \caption{Fit of the MWL SED of Mrk~421 observed in \cite{abdo_2011}, obtained wrapping the radiative provided by \agnpy within the classes for flux points handling and model fitting in \Gammapy. Figure from \cite{nigro_2022}, reproduced with permission of Astronomy and Astrophysics.}
    \label{fig:agnpy_mrk421}
\end{figure}

\agnpy \cite{nigro_2022} is a \python package modelling the radiative processes in jetted AGN. As \naima, \agnpy is entirely built in the \python scientific ecosystem and is one of the packages affiliated with the \astropy project. As for the other packages, \agnpy relies on a numerical approach to compute the radiative processes of non-thermal electron distributions. Routines for time evolution are not included in the package, though a module for the constraint of the spectral parameters according to a simple parametrisation of the acceleration and radiation processes is available. \agnpy implements synchrotron radiation, SSC and EC on anisotropic (accretion disk, BLR, DT) and isotropic (CMB) photon fields. Similarly to \gamera, the full angular dependency of the Compton cross section is taken into account, though only isotropic electron distributions can be considered. $\gamma\gamma$ absorption on all the photon fields target for Compton scattering can be computed (see e.g. Figure~\ref{fig:blr_absorption_agnpy}). Values for the opacity due to different EBL models \cite{franceschini_2008, finke_2010, dominguez_2011} are also included. The viewing angle $\theta_{s}$ of the observer to the jet axis is included among the parameters of all the physical processes implemented, such that \agnpy can be adopted to describe radio-loud AGN, beside blazars. In its current state it is not possible to consider multiple or complex emission regions beside the simple homogeneous sphere (blob). Utilities for data handling and fitting are not included in the package, a \Gammapy wrapper is provided instead. Living in the \python scientific ecosystem, \agnpy is seamlessly interfaceable with the fitting routines included in other tools such as \sherpa, as shown in several examples in the documentation and in \cite{nigro_2022}. In Figure~\ref{fig:agnpy_mrk421}, as an illustrative example, we show a fit of the MWL emission of Mrk421 obtained by wrapping \agnpy with \Gammapy. \agnpy is thoroughly validated by numerically checking the output of each radiative processes against results from the main reference used for its implementation \cite{dermer_2009a, dermer_2002, finke_2008, dermer_2009b, finke_2016} and against \jetset \cite[][see Sect. 4]{nigro_2022}. Examples of validation are illustrated in Figure~\ref{fig:blr_absorption_agnpy} (left panel) and in Figure~\ref{fig:validation}. Additional internal consistency checks are implemented: for example EC spectra and $\gamma\gamma$ opacities are compared against an approximation considering the target photon fields as a monochromatic point source (see e.g. Figure~\ref{fig:blr_absorption_agnpy}, right panel). Deviations well within $30\%$ are achieved when comparing against the literature and against \jetset, when the same physical assumptions are considered. Differences within a factor $2$ are instead obtained when comparing against processes implemented with different assumptions (e.g. when comparing against the EC implemented in \jetset). \agnpy is the first non-thermal modelling tools openly presenting such detailed numerical comparisons and integrating them in its test system. \agnpy has been used for modelling blazars, especially FSRQs thanks to its solutions for EC scattering and $\gamma\gamma$ absorption \cite{magic_2021_B1420, magic_2022_B0218, albert_2022}.
\par
\agnpy is hosted on \github\footnote{\url{https://github.com/cosimoNigro/agnpy}}, where 6 contributors are listed; its documentation is hosted on \readthedocs\footnote{\url{https://agnpy.readthedocs.io/en/latest/}}. The package includes a test suite part of its CI system. Numerical comparisons against literature reference and \jetset results are embedded in this tests. CD is also implemented, with each tagged version of the software made immediately available via \pip and \conda. \cite{nigro_2022} constitutes the release paper of the software.

\subsection{\bhjet}

\begin{figure}
    \begin{minipage}{0.48\textwidth}
        \centering
        \includegraphics[scale=0.35]{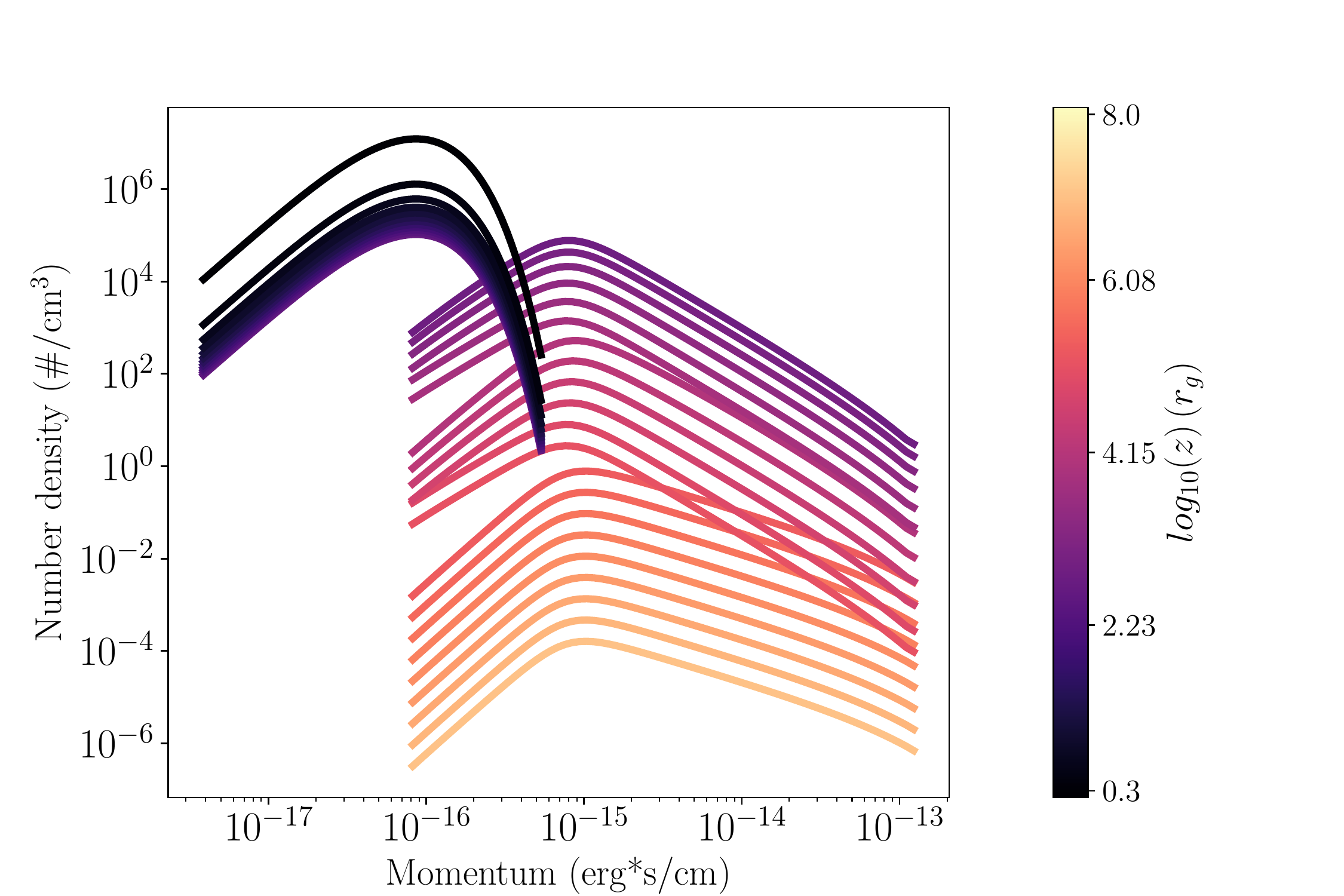}
    \end{minipage}\hfill
    \begin{minipage}{0.48\textwidth}
        \centering
        \includegraphics[scale=0.35]{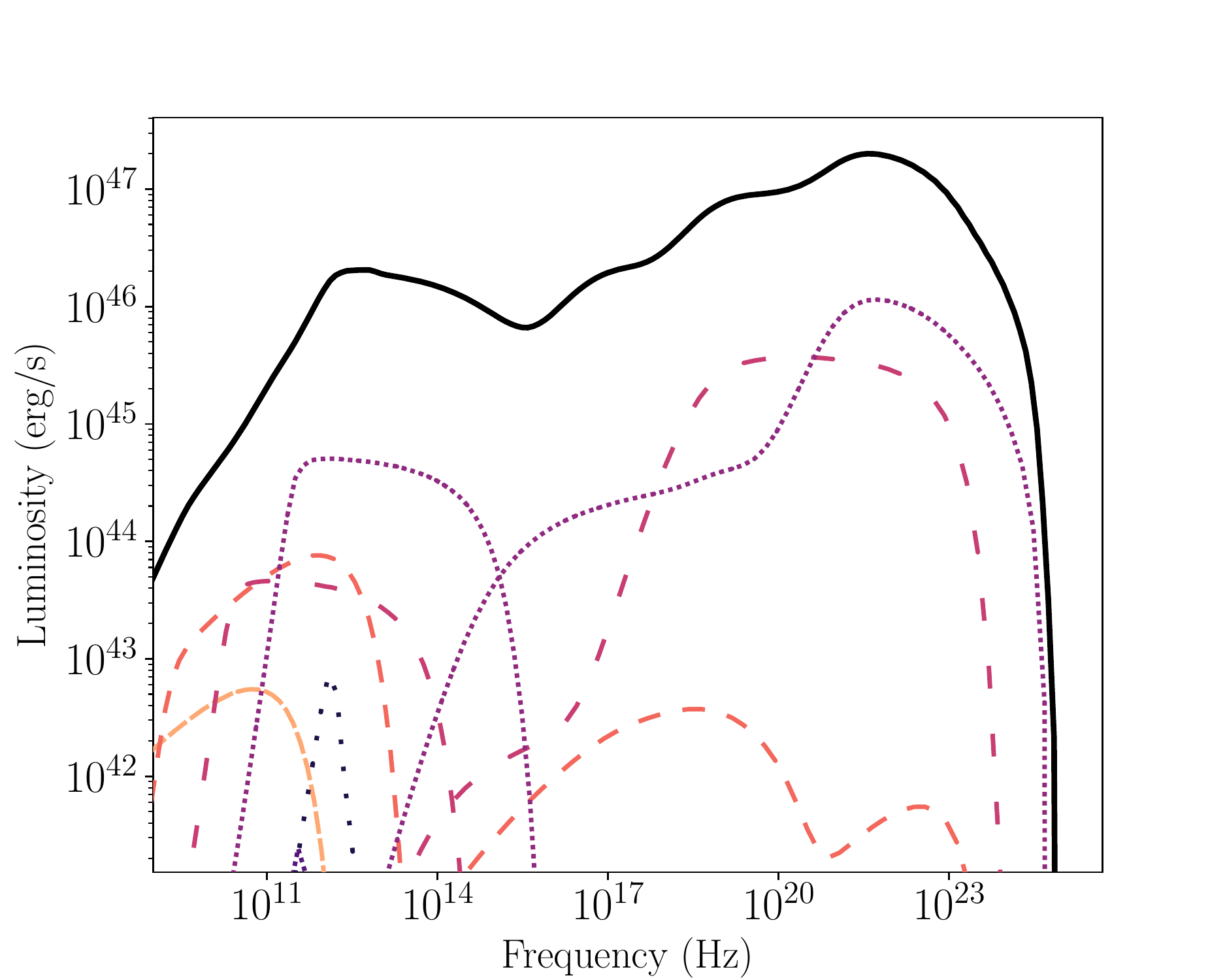}
    \end{minipage}
    \caption{Electron distributions and corresponding MWL SED computed with \bhjet at several distances along the jet axis of a FSRQ. The black line in the SED plot represents the sum of the emission at all heights. Figure from \cite{lucchini_2019}, reproduced with permission of the author.}
    \label{fig:fsrq_bhjet}
\end{figure}

\bhjet \cite{markoff_2001, markoff_2005, lucchini_2021} is a set of \CPP libraries modelling the emission of accretion / ejection systems of different scales: from black hole X-ray binaries (BHXB) to radio-loud AGN. It comprises the following libraries: \kariba, describing the radiative processes and their underlying particle distributions and the \agnjet and \bljet libraries, modelling jets with different physical properties. \agnjet describes a mildly relativistic, pressure-driven, jet \cite{falcke_biermann_1995, falcke_et_al_1995, falcke_biermann_1999} while \bljet describes a Blandford-K{\"o}nigl \cite{bk_1979}, magnetic-driven jet. A numerical approach is used for the radiative processes computations and a semi-analytical approach for the jet modelling calculations. As the library aims to describe also accretion systems, both thermal and non-thermal electron distributions can be considered. No time evolution is implemented, but a steady-state solution of the differential equation regulating the particle cooling can be evaluated, accounting for adiabatic, synchrotron and IC losses only in the Thomson regime. The radiative processes modelled by \kariba are: black body radiation, cyclotron radiation due to thermal electrons, synchrotron radiation due to non-thermal electrons, inverse Compton (SSC and EC on the AGN components). Successive scattering orders can be considered for the IC (i.e. the IC radiation can be target for further IC scattering). For both jet classes a fluidodynamic equation representing the velocity profile can be solved, allowing to obtain the particle density and the magnetic field at each height of the jet, and hence for the steady-state solution of the cooling calculated. \bhjet can therefore be used to evaluate the MWL emission from the entire outflow. No routines for SED fitting are provided but the array returned by the radiative processes computations is compatible with \texttt{XSPEC} \cite{arnaud_1996}. Some validation is provided for the radiative processes and the jet modelling in \cite{lucchini_2021}. The Compton spectra are benchmarked against calculations of the \texttt{compPS}\footnote{\url{https://heasarc.gsfc.nasa.gov/xanadu/xspec/manual/XSmodelCompps.html}} code, included in \xspec. The results of the semi-analytical modelling of the jet evolution (e.g. magnetic field, electron density) are instead compared against general-relativistic magnetohydrodynamics (GRMHD) simulations and it is observed that \bhjet can approximately reproduce the magnetic field value and the particle density in the case of a mildly-relativistic pressure-driven jet or in the case of a highly-relativistic magnetised jet. Before its public release, previous versions of the software have been extensively used for binaries and AGN modelling, \cite[see][and references therein]{lucchini_2021}. To illustrate an application of \bhjet to compute the spectrum of a VHE source, we show in Figure~\ref{fig:fsrq_bhjet} the electron distribution and the MWL SED of a FSRQ computed for several distances along the jet axis.
\par
\bhjet is available in \github\footnote{\url{https://github.com/matteolucchini1/BHJet}}, where 4 contributors are listed. No documentation is provided, but few example scripts to reproduce the results in \cite{lucchini_2021} are available. The library is not distributed with any package system and has to be manually downloaded from git and built with \make, along with the example scripts. \cite{lucchini_2021} constitutes the release paper.

\subsection{\flaremodel}

\begin{figure}
    \centering
    \includegraphics[scale=0.8]{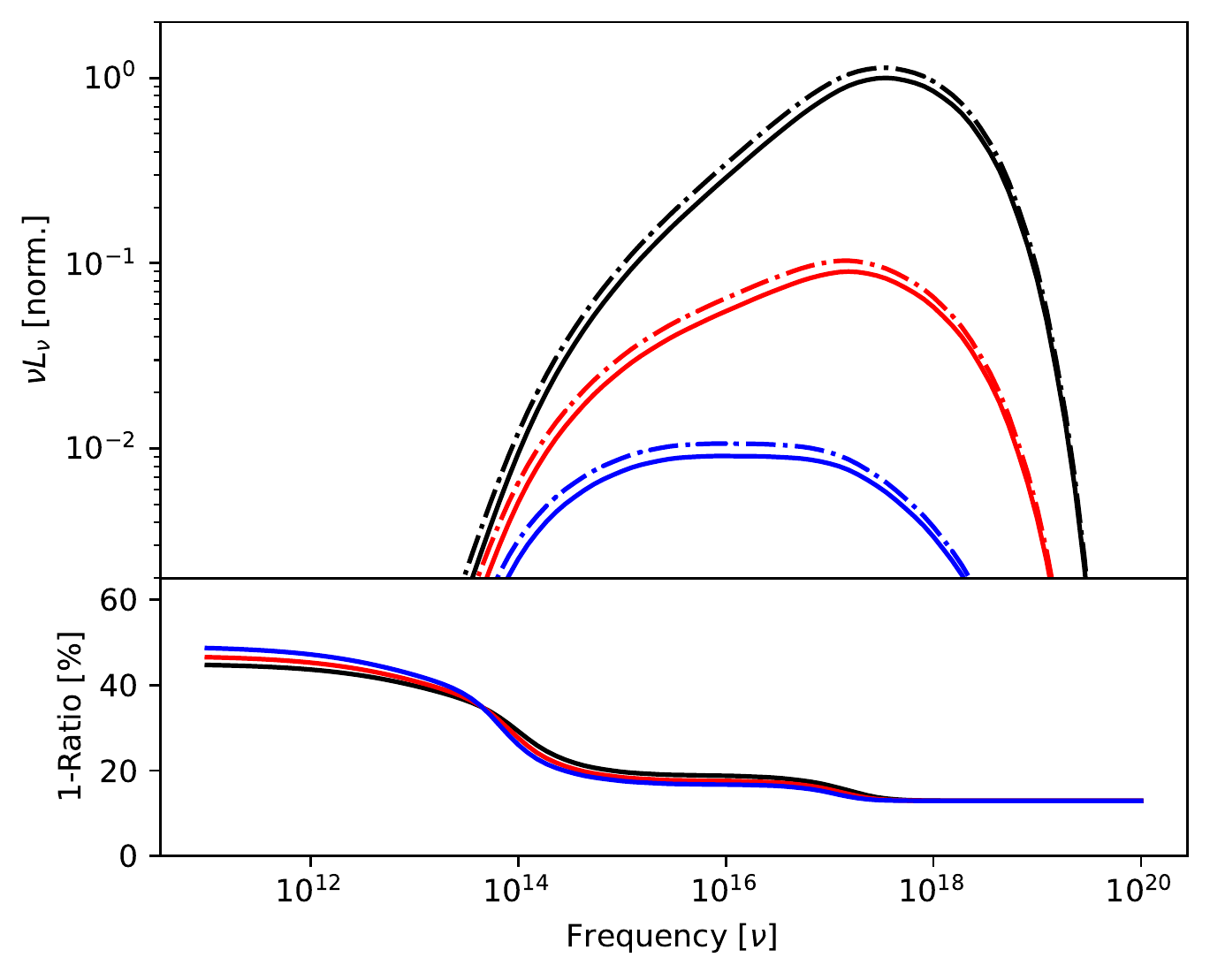}
    \caption{SSC emission computed with \flaremodel from a uniform emission region. SED with dashed line are obtained using the numerical ray-tracing approach, solid ones are obtained with the numerical integration considering a uniform particle distribution. Figure from \cite{dallilar_2022}.}
    \label{fig:ssc_flaremodel}
\end{figure}

\flaremodel \cite{dallilar_2022} is a python package modelling astrophysical synchrotron sources. Contrary to the uniform (spherical) emission regions considered in the other packages, \flaremodel allows to consider inhomogenous spherical emission regions. The basic routines are written in \CC with options for multi-threading and wrapped with a \python interface integrated with \numpy, but not with \astropy. Differently than the other codes, \flaremodel employs ray-tracing, i.e. the propagation of imaginary rays is followed through a region with changing physical conditions, hence allowing to consider non-isotropic particle distributions in the emission region. Both thermal and non-thermal electron distributions can be employed. Their time evolution is modelled considering adiabatic and synchrotron losses, while IC cooling is not included. Beside synchrotron radiation, also SSC emission can be computed. Multiple emission regions can not be considered. A SED fitting routine built on \texttt{lmfit} \cite{newville_2021} is made available. Validation is provided in \cite{dallilar_2022} for the computations implemented in the package. The synchrotron emissivity and absorption coefficients are compared against those computed with the \texttt{symphony} code \cite{pandya_2016}. The synchrotron emission obtained by \cite{band_1985} for a sphere with power-law radial density and magnetic field is reproduced. A consistency check, illustrated in Figure~\ref{fig:ssc_flaremodel}, is performed for the SSC from a uniform sphere, comparing the solution obtained with ray-tracing against the numerical simplification assuming a uniform particle distribution. Examples of time evolution under synchrotron and adiabatic cooling are provided in \cite{dallilar_2022}. The software was used in \cite{gravity_2021} to model the synchrotron emission of Sgr~A$^{\star}$.
\par
\flaremodel is hosted on \github\footnote{\url{https://github.com/ydallilar/flaremodel}}, where 1 contributor is listed. The documentation includes basic notebook tutorials\footnote{\url{https://flaremodel.readthedocs.io/en/latest/index.html}}, reproducing the figures in \cite{dallilar_2022}, constituting the release paper for the software. \flaremodel is distributed via \pip.

\section{Discussion and conclusions}
\label{sec:discussion}

\subsection{Review of the current packages}
We have examined in our review six packages modelling the non-thermal broad-band emission of jetted extragalactic sources from radio to gamma rays. We briefly consider also their capability to describe galactic sources characterised by the same emission mechanisms. In Table~\ref{tab:tools_physics_comparison} we present a global overview of the physical processes implemented by each software. All the tools provide leptonic synchrotron and SSC emission models. \bhjet and \flaremodel provide the most sophisticated calculations for these radiative processes, with the first taking into account the emission from the whole plasma outflow and several orders of IC scattering, and the latter employing ray-tracing to consider a non-uniform emission region. Having being used to model low-energy sources, \bhjet and \flaremodel are also the only libraries including thermal electron distributions. By considering the full angular dependency of the Compton cross section, \gamera can compute IC scattering with anisotropic electrons and anisotropic target radiation fields. Similarly, \agnpy provides IC scattering on anisotropic radiation field (with isotropic electrons though). Both \gamera and \agnpy can compute $\gamma\gamma$ absorption on anisotropic photon fields, while \naima and \jetset provide only absorption on the EBL. Regarding hadronic radiative processes, a description of the ${\rm p}{\rm p}$ interaction is implemented in \naima, \gamera and \jetset, while none of the tools includes photo-hadronic (${\rm p}\gamma$) emission models \cite{cerruti_2021}. In \naima and \gamera, only the $\pi_0$ decay in gamma rays is modelled, while \jetset models the decay of charged pions, computing the equilibrium distributions of secondaries $e^{\pm}$ pairs and their radiation (synchrotron, IC, and Bremsstrahlung). \jetset also computes the spectrum of $\nu$ produced in pion decays. For hadronic interactions both \gamera and \naima follow the parametrisation of \cite{kafexhiu_2014}, while \jetset implements the one of \cite{kelner_2006}. \gamera, \jetset and \flaremodel can numerically solve the differential equation describing the particles temporal evolution. Anyhow, while \flaremodel and \gamera take into account only the radiative cooling processes, \jetset provides also first and second order acceleration process, adiabatic expansion, and the possibility to have a decoupled radiative and acceleration regions. The remaining packages offer simplified alternatives: \naima allows for a custom particle distribution in input, \agnpy offers a constraint of the model parameters based on a simple parametrisation of the acceleration and radiation processes, and \bhjet provides the analytical solution of the differential equation at equilibrium. Concerning the type of sources that can be modelled, except for \naima, all the tools reviewed can be directly applied to describe the emission of jetted AGN. \naima, originally designed to model galactic high-energy sources, can be used to model jetted sources only through the manual implementation of the beaming pattern of the radiation. Incidentally, \naima was the only package used to model GRB emission. \bhjet and \jetset (through its plugins describing a microquasar or the blob expansion) are the only software considering the extended jet emission. The fit of a MWL SED with a tool ascribing the whole emission to a finite jet region underestimates the emission below a certain frequency in the radio band, as this is typically measured with a large integration region. As an example, one can see how the points below $10^{11}\,{\rm Hz}$ in Figure~\ref{fig:agnpy_mrk421} cannot be reproduced assuming synchrotron radiation from a blob, while these points are properly modelled in the microquasar model of \jetset, in Figure \ref{fig:jetset_fit}. We notice that the problem of integrating the non-thermal emission over a simple geometrical model of the jet has already been treated in the literature \cite{potter_2012, potter_2013a, potter_2013b, potter_2013c, tramacere_2022} and it would be an important implementation in the tools. Among the software considered, \bhjet is the only one suited to describe non-jetted low-power AGN (see the M81$^{\star}$ example in \cite{lucchini_2021}). None of the tools considered can be applied to describe other classes of extragalactic gamma-ray emitters, as starbust galaxies \cite{ohm_2016}.
\par
\begin{figure}
    \centering
    \includegraphics[scale=0.46]{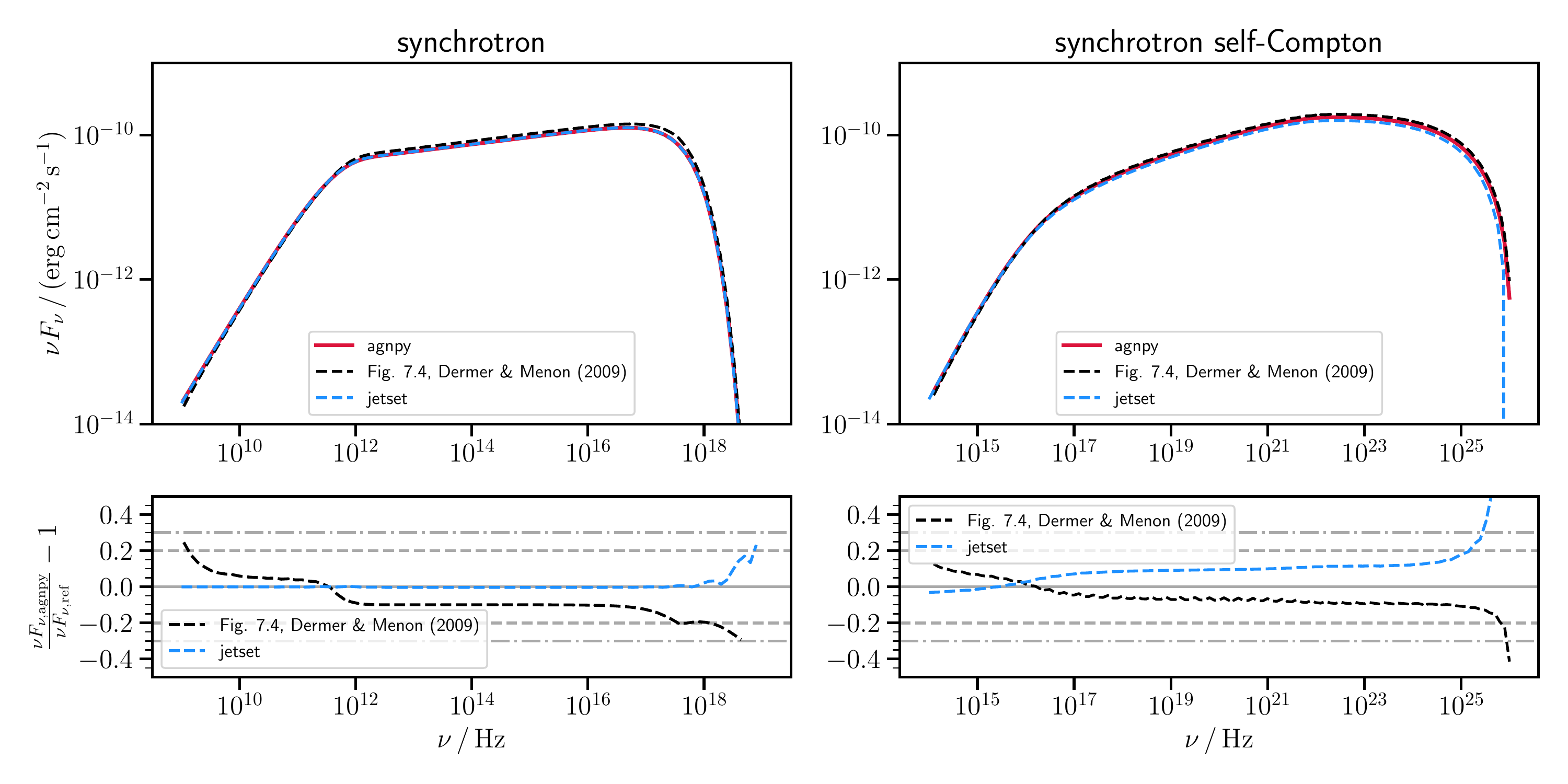}
    \caption{An example of validation: synchrotron and SSC SEDs generated with the same model parameters using \agnpy and \jetset. Both spectra are compared against a result from the literature \cite{dermer_2009a}. Figure from \cite{nigro_2022}, reproduced with permission of Astronomy and Astrophysics.}
    \label{fig:validation}
\end{figure}

In Table~\ref{tab:tools_software_comparison} we instead examine the compliance of individual tools with good modern software practices. We notice that \naima, \jetset, \agnpy, and \flaremodel are the ones simultaneously providing test suites, proper documentation and distribution via package managers. Though all the packages provide some degree of interface to fitting routines, \naima, \jetset, \agnpy, and \flaremodel, due to their interface with the \python scientific ecosystem, are also the ones better usable in combination with modern \python data-analysis tools and indeed provide wrappers to other data-analysis packages. We observe that in many of the software release papers, some degree of validation is provided, the most complete example being the cross-validation performed for \agnpy and \jetset in the release paper of the former \cite{nigro_2022}. Starting from the same set of model parameters, the SEDs obtained with the different software are compared against each other and against a reference SED from the literature (see Figure~\ref{fig:validation}). The other packages showed also a significant commitment to validation, with \bhjet benchmarking the Compton computation against the \texttt{compPS} code and the quantities obtained from the semi-analytical description of the jet evolution against GRMHD results. \flaremodel instead validated the synchrotron computations against the \texttt{symphony} software and and against the literature. Additionally, the SSC was internally checked comparing the ray-tracing and the numerical solutions (see Figure~\ref{fig:ssc_flaremodel}). Among the tools examined, \flaremodel is the only providing options for multi-threading (at \CC level). \jetset is instead the only framework with a specific class representing the model parameters. Having this object-oriented description of the parameters of the physical model allows one to impose physical limits, link them, and ultimately facilitates their wrapping with external packages (implementing their own parameters handling).

\subsection{Desiderata for future modelling packages}

The number and quality of software reviewed illustrates that the shift in paradigm towards an open-source modelling approach, described in the introduction, is already taking place. We might consider the tools covered by this review as a \textit{first generation} of open-source modelling tools. Therefore, after having reviewed the available software, we outline in this section what would be the \textit{desiderata} for the future generation of radiative modelling tools.

\begin{itemize}
    \item \textit{testing}: for such complex numerical models test suites are mandatory;
    \item \textit{validation}: this constitutes the most fundamental point. If a software has to be provided to a large community of astrophysicists, it is essential to provide a numerical validation against other software, or against reference templates. For example, benchmark SEDs, corresponding to a given physical scenario and a given set of model parameters, can be generated and shared as validation templates. This has already been proposed in \cite{nigro_2022};
    \item \textit{interfaceability}: as proposed by \cite{portegies_zwart_2018}, instead of several different packages one could envision a library of interfaceable \textit{fundamental solvers}, specialized, interconnectable, and respecting the single-responsibility principle. An example of combined workflow could be: obtaining the particle energy distribution as a result of the time evolution performed with one of these solvers, and then obtain the corresponding broad-band SED using the radiative processes of another solver. This would imply for the tools to be developed on a more fine-grained level, delegating the high-level interface to separate modules. Additionally, these basic blocks should have a minimal data / model interface, to facilitate the exchange of products. For example, particles distributions or radiative fields could use standardised specifications to interface with the classes handling them in the different solvers. Similarly, final products such as broad-band SEDs could be provided in the form of standardised (e.g. FITS \cite{wells_1981} or \astropy) tables with quantities (allowing units conversion). Table metadata could be used to store the model parameters (e.g.  parameters of the particle distribution, radius of the emission region, magnetic field intensity, etc.). Using standardised inputs and outputs, with a proper interface between the fundamental solvers, will make the validation process smooth and secure. High-level interfaces should finally orchestrate the fundamental solvers, linking the parameters of the basic blocks and facilitating the interface to other frameworks;
    \item \textit{data access}: as already demonstrated by the tools in this review, by living in the same computational ecosystem, modelling and data-analysis tools can be easily interfaced. The interface to specific analysis software, and eventually to online services providing astrophysical data, broadens the horizon of model fitting, allowing for example combination of data from different experiments, or from current and future generations of instruments. Moreover, having access to the instrument reduced data through the data-analysis packages would allow to perform a more accurate fit of the physical model, for example folding it with the instrument IRF and computing a Poissonian likelihood of the observed and expected counts (as commonly done in X-ray and gamma-ray astronomy \cite{piron_2001} with simple analytical models, PL, log-parabola, etc.). Due to the current limitations, a $\chi^2$ fit is commonly performed to flux points that are often computed making assumption on the underlying shape of the photon spectrum and never provided with a matrix quantifying their correlations. \naima, \agnpy, and \jetset already demonstrate the possibility of forward-fold fit of high-energy astrophysical data through their \sherpa and \Gammapy wrappers\footnote{Both \sherpa and \Gammapy can read the PHA OGIP standard adopted to represents counts and IRF of X-ray (\url{https://heasarc.gsfc.nasa.gov/docs/heasarc/ofwg/docs/spectra/ogip_92_007/node5.html}) and Gamma-ray (\url{https://gamma-astro-data-formats.readthedocs.io/en/latest/spectra/ogip/index.html}) instruments.};
    \item \textit{accessibility}: we remark that making the code available online with a license it is not sufficient to make it properly accessible. Care has to be taken by developers to write a proper documentation. The latter not only serves didactic purposes, for example it can be easily used in hands-on tutorials, but plays the fundamental role of forming future users or developers.
\end{itemize}

\subsection{Conclusions}

In this review we have presented the state-of-the-art of open source, reproducible, frameworks modelling the radiative processes in extragalactic VHE gamma-ray sources. We have highlighted the main features of the presented packages in terms of the physical processes they implement, and in terms of the good software practices they comply with. We consider these packages as a first generation of software paving the road for a future generation of frameworks realising a fully-reproducible modelling of high-energy astrophysical sources. Before concluding, broadening the scope of our review, we would like to remark the following points concerning physical interpretation:
\begin{enumerate}
    \item The current packages represent mostly single- or few-developers projects, with a strong commitment and effort from few individuals, who are offering a scientific product to the community, fulfilling the full chain from coding, to documentation and distribution;
    \item despite the aforementioned efforts, the attitude to publish scientific article based on accessible and reproducible models is not yet standard. Closed-source software, if used in scientific publications, should at minimum be accessible in form of binaries, to allow the astrophysical community to reproduce and validate what has been published;
    \item even though the presented products reach high-quality standards, none of them covers the entire panoply of physical processes, and a large overlap of features among the products is present. In this sense, the most desirable solution would be an effort to produce a library of interfaceable fundamental solvers, with a strong support from the community, the large collaborations, and the editorial boards.
\end{enumerate}

\startlandscape
\begin{table}[H] 
\caption{Physical processes implemented in the software reviewed.}
    \begin{tabularx}{\textwidth}{|l|l|l|l|l|l|l|l|l|l|l|l|l|l|}
    \toprule
    \textbf{software} & \textbf{sources} & \textbf{approach} & \multicolumn{3}{c|}{\textbf{particles}}                      & \multicolumn{6}{c|}{\textbf{processes}}                                                                         & \textbf{temp. ev.} & \textbf{emission region}\\
                      &                  &                   & \textbf{thermal} & \multicolumn{2}{c|}{\textbf{non-thermal}} & \multicolumn{4}{c|}{\textbf{leptonic}}                              & \textbf{hadronic}  & \textbf{absorption}  &                    &    \\
                      &                  &                   &                  & ${\rm e^{\pm}}$  & ${\rm p}$              & \textbf{synch.} & \textbf{SSC} & \textbf{EC}      &\textbf{Brems.}  & \textbf{pp}        & $\gamma\gamma$       &                    &     \\
    \midrule
    \naima            & PWN, SNR, GRB    & numerical         & \xmark           & \cmark           & \cmark                 & \cmark          & \cmark       & \cmark (CMB)     & \cmark         & \cmark$^{\dagger}$  & \cmark (EBL)        & \xmark              & not specified     \\ \hline
    \gamera           & PWN, SNR, AGN    & numerical         & \xmark           & \cmark           & \cmark                 & \cmark          & \cmark       & \cmark$^{\odot}$     & \cmark         & \cmark$^{\dagger}$  & \cmark$^{\star}$        & \cmark              & multiple uniform  \\
                      & microquasars     &                   &                  &                  &                        &                 &              &                  &                &                     &                     & (only cool.)        &                   \\ \hline
    \jetset           & jetted AGN, PWN  & numerical         & \xmark           & \cmark           & \cmark                 & \cmark          & \cmark       & \cmark           & \cmark         & \cmark$^{\ddagger}$ & \cmark (EBL)        & \cmark              & multiple uniform  \\
                      & microquasars, SNR &                   &                  &                  &                        &                 &              &                  &                &                     &                     & (acc. + cool.)      & acc. + rad.       \\ \hline
    \agnpy            & jetted AGN       & numerical         & \xmark           & \cmark           & \xmark                 & \cmark          & \cmark       & \cmark$^{\star}$ & \xmark         & \xmark              & \cmark$^{\star}$    & \xmark              & single uniform    \\ \hline
    \bhjet            & binaries, AGN    & numerical         & \cmark           & \cmark           & \xmark                 & \cmark          & \cmark       & \cmark           & \xmark         & \xmark              & \xmark              & \xmark              & whole jet         \\
                      &                  & semi-analytical   &                  &                  &                        &                 &              &                  &                &                     &                     &                     &                   \\ \hline
    \flaremodel       & synch. sources   & numerical         & \cmark           & \cmark           & \xmark                 & \cmark          & \cmark       & \xmark           & \xmark         & \xmark              & \xmark              & \cmark              & single            \\
                      &                  & ray-tracing       &                  &                  &                        &                 &              &                  &                &                     &                     & (only cool.)             & radial dep.       \\ \hline
    \bottomrule
    \end{tabularx}
    \begin{adjustwidth}{+\extralength}{0cm}
        \noindent\footnotesize{\textsuperscript{$\dagger$} ${\rm p}{\rm p}$ interaction: computing only gammas from $\pi_0$ decay.}
        \newline 
        \noindent\footnotesize{\textsuperscript{$\ddagger$} ${\rm p}{\rm p}$ interaction: computation of radiation from secondaries of charged pions (pairs evolved in time to equilibrium) and of $\nu$ spectra.}
        \newline
        \noindent\footnotesize{\textsuperscript{$\odot$} full angular dependency of the Compton cross section: anisotropic electrons and anisotropic photon fields.}
        \newline
        \noindent\footnotesize{\textsuperscript{$\star$} full angular dependency of the Compton or $\gamma\gamma$ cross sections: anisotropic photon fileds.}
    \end{adjustwidth}
    \label{tab:tools_physics_comparison}
\end{table}
\finishlandscape

\begin{table}[H]
\caption{Compliance of software reviewed with modern good software practices.}
	\begin{adjustwidth}{-\extralength}{0cm}
		\newcolumntype{C}{>{\centering\arraybackslash}X}
		\begin{tabularx}{\fulllength}{CCCCCCC}
			\toprule
			\textbf{software} & \textbf{language} & \textbf{license} & \textbf{documentation} & \textbf{installation} & \textbf{CI or test units} & \textbf{CD}\\
			\midrule
			\naima	& \python & BSD-3$^1$ & Read the Docs & \pip, \conda & yes & no\\
			\gamera & \CPP, \python & not specified & GitHub Pages & \make file & minimal & no \\
			\jetset & \CC, \python & BSD-3 & Read the Docs & \pip, \conda & yes & yes\\
			\agnpy & \python & BSD-3 & Read the Docs & \pip, \conda & yes & yes \\
			\bhjet & \CPP & MIT$^2$ & no & \make file & no & no \\
			\flaremodel & \CC, python & BSD-3 & Read The Docs & \pip & yes & no \\
			\bottomrule
		\end{tabularx}
	\noindent\footnotesize{\textsuperscript{$1$} \url{https://opensource.org/licenses/BSD-3-Clause}}
    \newline 
    \noindent\footnotesize{\textsuperscript{$2$} \url{https://opensource.org/licenses/MIT}}
    \end{adjustwidth}
    \label{tab:tools_software_comparison}
\end{table}

\vspace{6pt} 


\funding{This work was supported by the European Commission's Horizon 2020 Program under grant agreement 824064 (ESCAPE—European Science Cluster of Astronomy and Particle Physics ESFRI Research Infrastructures), by the the ERDF under the Spanish Ministerio de Ciencia e Innovaci\'on (MICINN, grant PID2019-107847RB-C41), and from the CERCA program of the Generalitat de Catalunya.}


\conflictsofinterest{The authors are the main developers of the  \agnpy (C. Nigro) and \jetset (A. Tramacere) packages described in the review.} 

\abbreviations{Abbreviations}{
The following abbreviations are used in this manuscript:\\

\noindent 
\begin{tabular}{@{}ll}
AGN & Active Galactic Nucleus \\
GRB & Gamma-ray Burst \\
MWL & multi-wavelength \\
CTA & Cherenkov Telescope Array \\
BH & black hole \\
IR & infrared \\
FSRQ & Flat Spectrum Radio Quasars \\
PL & power law \\
SED & spectral energy distribution \\
IC & inverse Compton \\
SSC & synchrotron self-Compton \\
EC & external Compton \\
BLR & broad line region \\
DT & dust torus \\
CMB & cosmic microwave background \\
EBL & extragalactic background light \\
MCMC & Markov chain Monte Carlo \\
PWN & pulsar wind nebula \\
ISM & interstellar medium \\
IRF & instrument response function \\
BBB & big blue bump \\
CI & continuous integration \\
CD & continuous deployment \\
BHXB & black hole X-ray binary \\
GRMHD & general-relativistic magnetohydrodynamics \\
\end{tabular}}




\begin{adjustwidth}{-\extralength}{0cm}

\reftitle{References}

\end{adjustwidth}
\end{document}